\documentclass[preprint]{aastex}
\usepackage{natbib}
\usepackage{graphicx}
 \usepackage{color}

\received{--}
\revised{--}
\accepted{--}
\cpright{}{2010}
\slugcomment{To be submitted to ApJ}
\shorttitle{LASCO CME Statistics}
\shortauthors{Vourlidas et al.}
\begin{document}

\title{Comprehensive Analysis of Coronal Mass Ejection Mass and Energy
  Properties Over a Full Solar Cycle}

\author{A.Vourlidas\altaffilmark{1}, R. A. Howard\altaffilmark{1},
E. Esfandiari\altaffilmark{2}, S. Patsourakos\altaffilmark{3},  S. Yashiro\altaffilmark{4}, and G. Michalek\altaffilmark{5}}  

\altaffiltext{1}{code 7663, Naval Research Laboratory, Washington, DC, USA.}

\altaffiltext{2}{Adnet Systems Inc, Rockville, MD, USA}

\altaffiltext{3}{University of Ioannina, Department of Physics,
  Section of Astrogeophysics, Ioannina, Greece} 

\altaffiltext{4}{Center for Solar and Space Weather, Catholic
  University of America, Washington, DC, USA} 

\altaffiltext{5}{Astronomical Observatory of Jagiellonian University,
  Cracow, Poland}
%
%
\begin{abstract}
  The LASCO coronagraphs, in continuous operation since 1995, have
  observed the evolution of the solar corona and coronal mass
  ejections (CMEs) over a full solar cycle with high quality images
  and regular cadence. This is the first time that such a dataset
  becomes available and constitutes a unique resource for the study of
  CMEs. In this paper, we present a comprehensive investigation of the
  solar cycle dependence on the CME mass and energy over a full solar
  cycle (1996-2009) including the first in-depth discussion of the
  mass and energy analysis methods and their associated errors. Our
  analysis provides several results worthy of further studies. It
  demonstrates the possible existence of two event classes; 'normal'
  CMEs reaching constant mass for $>10$ R$_{\sun}$ and 'pseudo' CMEs
  which disappear in the C3 FOV. It shows that the mass and energy
  properties of CME reach constant levels, and therefore should be
  measured, only above $\sim 10 R_\sun$. The mass density
  ($g/R_\sun^2$) of CMEs varies relatively little ($<$ order of
  magnitude) suggesting that the majority of the mass originates from
  a small range in coronal heights. We find a sudden reduction in the
  CME mass in mid-2003 which may be related to a change in the
  electron content of the large scale corona and we uncover the
  presence of a six-month periodicity in the ejected mass from 2003
  onwards.
\end{abstract}

\keywords{Sun: activity -- Sun: corona -- Sun: coronal mass ejections
  -- Sun: magnetic fields}
\section{Introduction}

Coronal Mass Ejections (CMEs) have been observed since the 1970s but
in the last 15 years they have moved to the forefront of coronal
research as evidenced by hundreds of publications. This large amount
of work is driven by the availability of continuous solar observations
from the instrumentation aboard the SOHO mission \citep{soho:95}. The
Large Angle and Spectroscopic Coronagraph (LASCO) suite
\citep{lasco:95}, in particular, has revolutionized CME analysis with
high cadence, high spatial resolution and high sensitivity
observations of these events. Since the early phases of the mission, a
concerted effort was initiated to generate a comprehensive event list
based on visual inspection of the images and manual measurements of
several CME properties such as speed, width, and position angle. The
LASCO CME list grew to 14,000 events by mid-2009 and is the largest
collection of CME properties readily available online. This resource
has been used extensively to study various CME properties
\citep{2000JGR...10518169S, 2004JGRA..10907105Y,
  2009EM&P..104..295G}. In 2008, the LASCO instruments reached another
milestone. They provided the first ever CME database to cover a full
solar cycle with a single instrument, thus opening investigations to
the solar cycle dependence of CME properties
\citep{2004AdSpR..34..391G,2007AdSpR..40.1042C,mittal_properties_2009}
without having to worry about inter-instrument differences.  Most of
the previous statistical analyses have focused on the properties of
speed, width, rate of occurrence and association with flares and other
low coronal phenomena. These properties are most commonly used in CME
research.  In 2004, we added CME mass and kinetic energy measurements
to the online list and they have been incorporated in some analyses
\citep{2009EM&P..104..295G,mittal_properties_2009}. However, CME mass
and energy measurements are used rather infrequently in analysis of
individual events, most likely because researchers are not familiar
with the mass analysis methods. There has not been a comprehensive
discussion of the procedures, assumptions, errors and limitations of
CME mass and energy analysis except for some descriptions in our
papers \citep{vour:00, vour:02, prasad:07} and a few past works
\citep{1975SoPh...42..163H,1981SoPh...69..169P}. The present paper
aims to fill this gap in the CME literature (Sections~1-2). In Section
3, we describe the database of CME mass images and the
associated mass and energy statistics. Only a small part of this
information is included in the online LASCO CME list. While the mass and energy
statistics have been briefly touched on in past studies, section 4
presents a detailed statistical analysis of the mass and energy
properties expanding on the previous work of \citet{russ:85} and
\cite{vour:02}. We discuss several aspects that emerge
from the statistical analysis of such a large event sample such as
solar cycle dependencies and the possible existence of
'pseudo-CMEs'. We conclude in \S~5.

\section{Calculation of CME Mass}

The white light emission of CMEs, and of the quiescent corona, arises
from the scattering of photospheric light by coronal
electrons. The emission is optically thin and therefore contains
contributions from all electrons along a given line of sight (LOS). It
is also linearly polarized. The emission mechanism, Thomson
scattering, is well understood and the total and polarized brightness
intensities can be easily calculated if the LOS electron distribution
is known or can be estimated \citep{billings:66, hayes:01,
  v_h:06}. Inversely, if the emission of a coronal feature is measured,
then the number of electrons (and hence mass) can be estimated.

To measure the mass of a CME, we first need to separate the brightness
contribution of the CME from the background coronal signal. Given the
dynamic nature of the ejection, the best way to remove the background
corona is to subtract a suitably chosen pre-event image from the image
containing the CME. This has been the procedure for all mass measurements
published to date
\citep[e.g.,][]{stewart_mccabe_koomen_hansen_dulk_1974,
  1981SoPh...69..169P, russ:85, 1994ESASP.373..409H, vour:02} and has
been described before \citep{1994ESASP.373..409H, vour:00,vour:02}. In this paper, we provide a
thorough description of the procedures for mass and energy analysis
starting with an outline of the steps involved:

\begin{enumerate}
\item The time series of the CME are examined to select the CME images
  and a pre-event image that does not contain signatures of the CME or
  other disturbances (e.g., a previous CME) over the range of position
  angles occupied by the event in question. The pre-event image must
  also be close in time to the event to minimize solar rotation or
  evolutionary effects. We will return to these issues in the next
  section.
\item The CME and pre-event images are corrected for instrumental
  effects (e.g., flat-fielding, vignetting, exposure time variation, stray light,
  etc) and calibrated in mean solar brightness (MSB or B/B$_{\sun}$)
  units to produce the so-called Level-1 images\footnote{Level-1
    images are available online at the NRL LASCO webpage
    (\url{http://lasco-www.nrl.navy.mil})}.

\item The pre-event image is subtracted from the images containing the
  CME. The resulting images are now in units of excess
  MSB and correspond to a base-difference sequence,
  roughly speaking.
\item The excess brightness image is transformed to the number of
  excess electrons using the Thomson scattering equations
  \citep{billings:66} and the assumption that all electrons lie on the
  plane of the sky (Section~2.1.4 for details). This assumption is
  driven by the unknown LOS distribution of the CME electrons and
  provides a convenient lower limit to their number.
\item The mass (per pixel) can then be calculated from the number of
  electrons by assuming a composition of 90\% H and 10\% He which corresponds to
  $1.97\times 10^{-24}$ g per electron \citep{1975SoPh...42..163H}.
\item Finally, the CME mass is calculated by summing all pixels inside
  the CME region. The region can be defined in various ways depending
  on the objective. Most often, the observer outlines the CME manually
  as in \citet{vour:00} and \citet{prasad:07}. We call this the
  'region of interest (ROI) method'. Alternatively, one can define a
  sector based on the CME width and central position angle and the
  coronagraph occulter edge and height of the CME front. These
  observed parameters define the angular boundaries and the inner and
  outer radial boundaries, respectively. We call this the 'sector
  method' and is the method used for the measurements in this paper. A
  third option is to measure mass flow though a location in the corona
  by defining a narrow (i.e., a few pixels wide) rectangular box, and
  position it tangent to the limb, at a given height. This setup
  emulates the observations of slit-based instruments, such as coronal
  spectrometers, and can be used to compare coronagraphic and
  spectroscopic observations as we have done in
  \citet{2003ApJ...597.1118C}.
\end{enumerate}
It is obvious that a number of assumptions enter in the final mass
calculation. A top level assessment of the reliability of CME mass
measurements, based on Thomson scattering properties \citep{vour:00}
and MHD simulations \citep{2005ApJ...627.1019L}, has shown that the
masses may be underestimated by $\sim 2$x.
\subsection{Error Analysis of CME Mass Calculations}
The calculation of the mass of a CME involves several analysis steps;
the images have to be calibrated, subtracted and their intensities
converted to mass values. Errors and assumptions are associated
with every step. However, this procedure has never been described in
detail before and several of the steps are general and applicable to
other studies besides mass analysis. We use an actual example to
structure the discussion. 

\begin{figure} 
\includegraphics[width=3in]{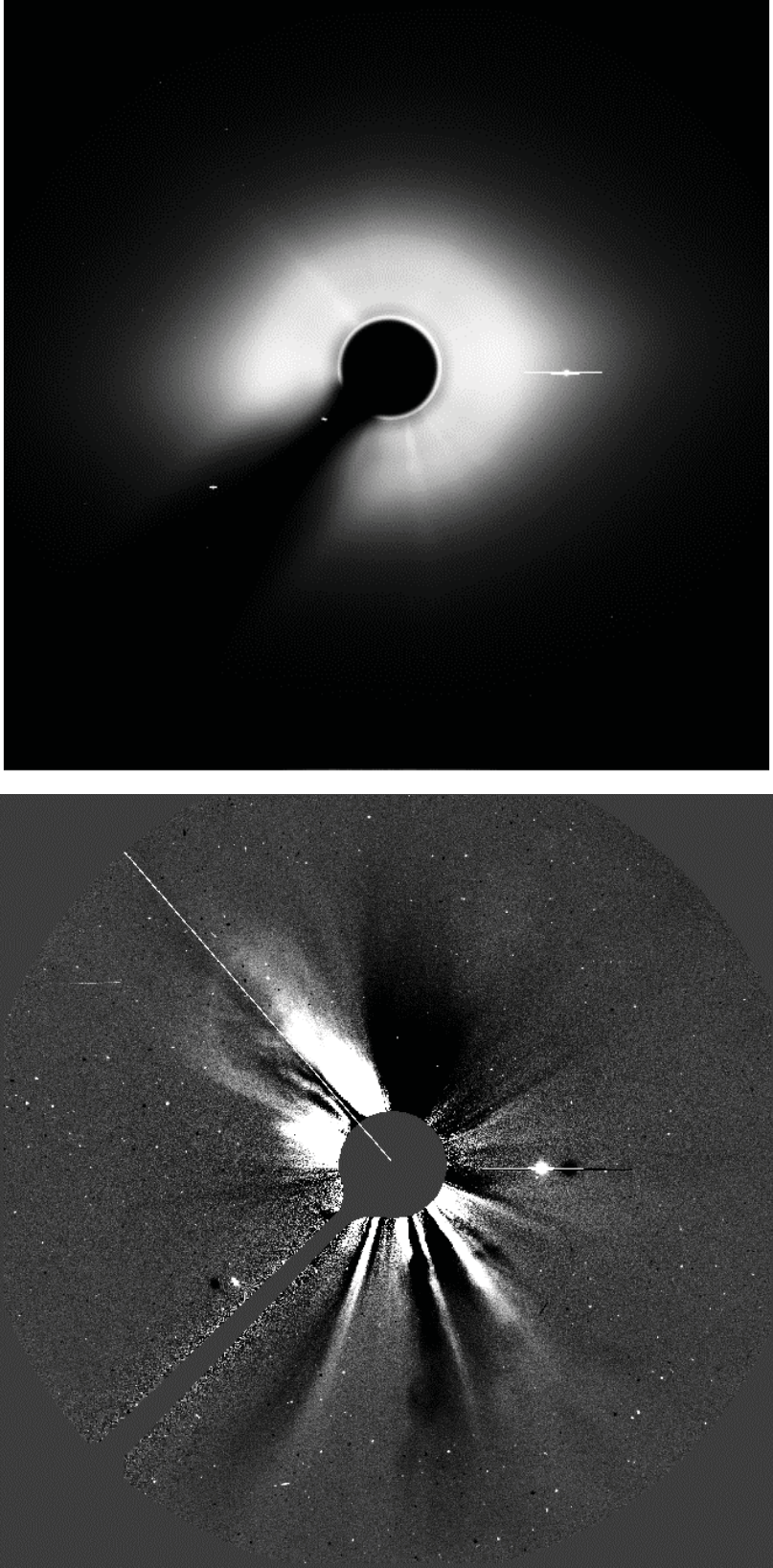}
\caption{\textsl{Top panel:\/} A typical LASCO/C3 image of a CME as
  received on the ground. \textsl{Bottom panel:\/} An excess mass
  image prpduced by the image above after removal of the F-corona,
  instrumental effects and the subtraction of a pre-event image. The
  solid line marks the position angle of the radial brightness profile we use in our analysis.  }\label{fig:example}
\end{figure}

We transform a C3 image, containing a CME snapshot, from its raw stage
to a mass image (Figure~\ref{fig:example}). We chose C3 rather than C2
images because the C3 instrument has a larger field of view (FOV),
CMEs are fainter and the influence of non-solar sources such as
F-corona, stars and cosmic rays is larger. Therefore, errors in C3
masses are expected to be higher than in C2 and our error analysis
provides a conservative upper bound. For the same reason, we have
chosen a rather weak CME event occurring in the presence of bright
background structures (i.e., streamers) which reflects the majority of
cases in the LASCO database. The procedure to achieve the
transformation from raw images to mass can be outlined as follows:
\begin{displaymath}
{{DN}\over{sec}} \rightarrow \textrm{MSB} \rightarrow \textrm{Excess MSB} \rightarrow \textrm{No
of e}^- \rightarrow {g} 
\end{displaymath}
Let us take a look at each one of these steps in detail. 
\subsubsection{From DN to Brightness}
After the telemetry data are received on the ground, they are unpacked,
and reformatted by the LASCO pipeline
and are stored in the database as Level-0.5 FITS files in units of Digital Numbers (DNs). The first step
towards calculating CME masses is to calibrate the images \citep{jeff:06,2006Icar..182..281L} to
Level-1 or in other words, we convert them from DNs to brightness
units (per pixel) as follows
\begin{equation} 
I = {{S}\over{t}}*V*F_{cal} \label{eq:dn2msb}  
\end{equation} 
where $I$ is the brightness in mean solar brightness units (MSB), $S$
the signal in the image (DN), $t$ the exposure time in seconds, $V$
the vignetting function, and $F_{cal}$ is the calibration factor
(MSB/DN/sec/pix) for the passband of the instrument. Note that there
is a different calibration factor for each filter. MSB is a standard
quantity in coronagraph data analysis and represents the ratio of the
coronal brightness ($B$) at a given pixel location to the average
brightness of the solar disk ($B_{\sun}$). Sometimes it is denoted as
$B/B_{\sun}$ but we will maintain the MSB notation here. The
vignetting function is a property of the optical system measuring the
amount of obscuration of the aperture due to the occulters and other
stops, as a function of field angle. Its effect on the image
corresponds to that of a radial density filter combined with spatial
resolution degradation in the inner field of view. The calibration
factor is initially determined on the ground and refined in-orbit
through the analysis of stellar photometry. Details on these and other
coronagraph calibrations can be found in \citet{jeff:06,2006Icar..182..281L}.
Equation~(\ref{eq:dn2msb}) is a rather simplified description of the
image calibration retaining only terms important for our
discussion. The actual Level-1 calibration procedure implements a host
of other corrections (e.g., stray light subtraction, roll to solar
north, etc) which are of secondary importance here.

For the error analysis, we use the ratio of the standard deviation,
$\sigma_x$, of a given quantity, $x$, over the quantity itself
($\sigma_x/x$) and perform standard error propagation calculation to
assign the final error. For Equation~(\ref{eq:dn2msb}), the results are
collected in Table~1. The signal in the image (when converted to photons)  obeys
photon statistics and is given by $1/\sqrt{S}$. Obviously, the error
increases as a function of heliocentric distance as the CME gets
fainter.  The other errors have been discussed in \citet{jeff:06}.
When we apply these error estimates in Equation~(\ref{eq:dn2msb}), we
find ${{\sigma_I}\over{I}} \approx 1.98\% - 2.3\%$. In other words, the
photometric accuracy of a single C3 image is extremely good, of the
order of a few percent.

\subsubsection{From MSB to Excess MSB} 
\begin{figure} 
\includegraphics[width=3in]{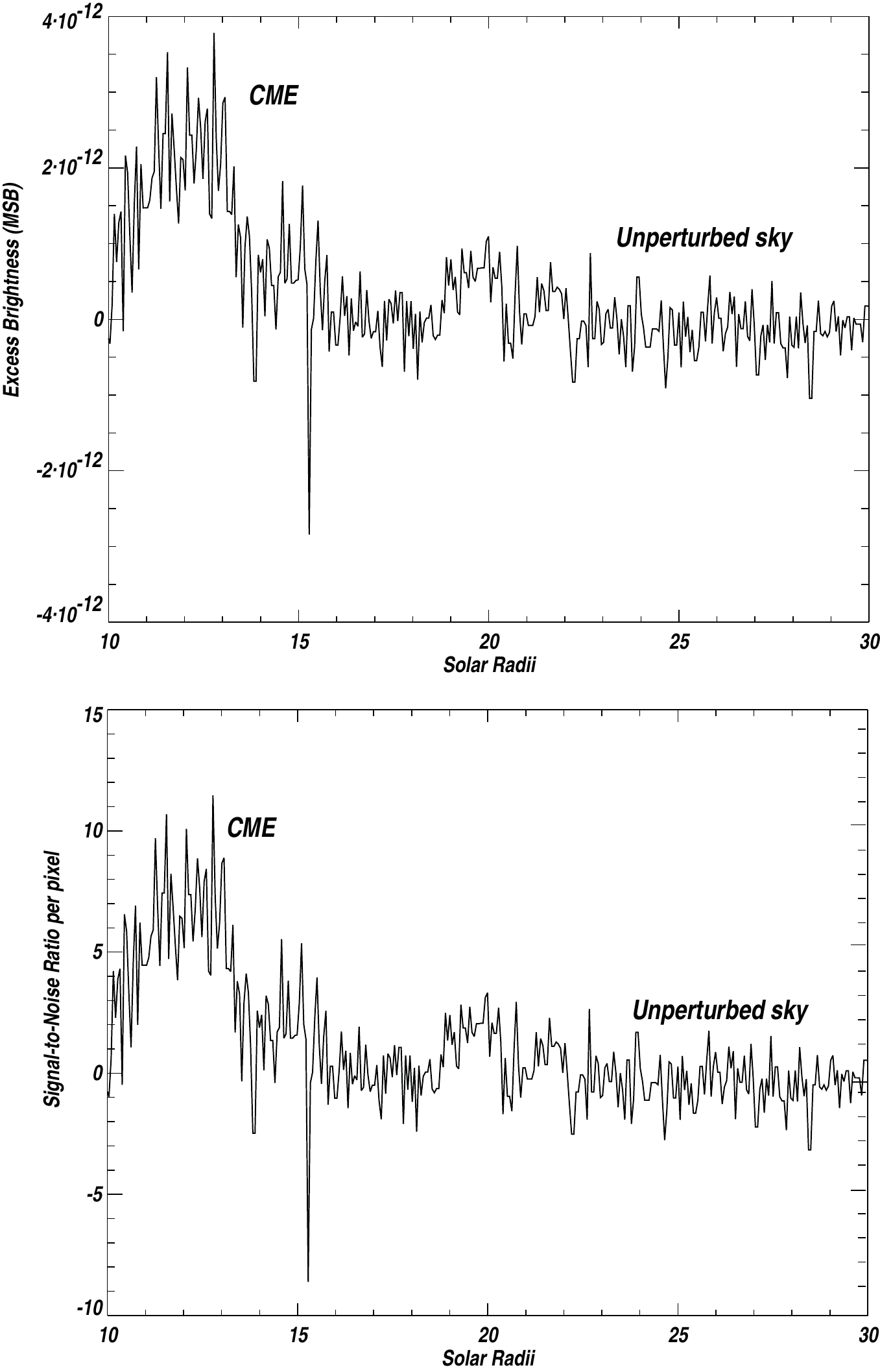}
\caption{\textsl{Top panel:\/} A typical radial profile of excess
  brightness. The CME and background sky levels are shown. The units
  are in MSB per pixel. \textsl{Bottom panel:} The same plot in units of
  SNR per pixel. The position angle of the brightness profile is shown
  in Figure~\ref{fig:example}. } \label{fig:excess_stat}
\end{figure}

Although the individual Level-1 images are photometrically accurate
and without instrumental defects, they are still difficult to use for
analysis because the F-corona, the background streamer structure, and any remaining stray light component hinder
the accurate detection of the faint CME boundaries. The F-corona
becomes increasingly the dominant contributor to the observed signal
above about 5-6 R$_\sun$ while streamers tend to affect the
visibility in the inner corona. The most straightforward way to remove
these effects is by subtracting an image which does not contain any
CME signal. This pre-event image must be as close in time as possible
to the CME to minimize effects from coronal evolution and should not
contain brightness changes due to CME material (current or a previous
CME) or CME-induced effects, such as streamer deflections. Any remaining stray light component will be completely removed as well. These
requirements are not always easy to satisfy especially during active
periods when up to 10 CMEs may erupt in a single day. Sometimes, a
post-event image can be used, but this is more difficult to put into
an automatic routine, because it is difficult to automatically determine
the end-point of the CME event. In practice, the user
has to ensure that the pre-event image does not contain any defects
or effects from other events only over the position angles covered by
the CME in question. The rest of the image can contain other events or
even missing telemetry blocks. The resulting image is referred to as an 'excess
brightness' image and is given by
\begin{equation} 
I_{CME} = I - I_{pre}, \quad \sigma_I/I \approx 4\% 
\end{equation} 
where $I$ and $I_{pre}$ are the CME and pre-event
image, respectively, and the error is propagated from the errors found
in the previous section. The error is based on the implicit assumption that the errors in the
two images are statistically independent. This is not a good
assumption, however. There is a high degree of correlation
between images since the location of streamers and even the
position angles of other CMEs may be similar in the two images. It is
very difficult to provide an analytical error estimate given the
multitude of event sizes, and coronal topologies. Instead we use the actual signal statistics from the excess brightness image
in our example. Figure~\ref{fig:excess_stat} shows a radial profile
across the CME and the undisturbed background ahead of the event
labeled as 'sky background'. The signal levels are $1.3\pm 3.4\times10^{-13}$ MSB. In our experience, these levels (of the
order of $10^{-13} MSB$) are very typical for C3 images. A more
intuitive representation is in terms of SNR per pixel shown in the
bottom panel of Figure~\ref{fig:excess_stat}. The CME is about 5-10
times brighter than the background in a running difference
image. Again, this is a very typical value for an average CME
event. We can calculate the error based on these numbers but this is
again an overestimation. Rarely, if ever, does one need to calculate
the CME mass for a single pixel. We almost always sum the mass
(brightness in this case) over hundreds or thousands of pixels. In our
example, the CME occupies $\sim38,000$ pixels resulting in total
brightness of $3.7\times10^{-8}\pm 2\times10^{-12}$ MSB. The
corresponding values for the background are $-1.7\times10^{-9}\pm
5\times10^{-13}$ MSB. The
summation naturally minimizes the errors. We summarize the various
methods and errors in Table~1. It is rather
obvious that the subtraction of a pre-event image introduces
insignificant errors, if the pre-event image is selected with care.

\subsubsection{From Excess MSB to Excess Mass} \label{sec:mass} At
this stage, we have isolated the brightness due to the CME material
from that of the background corona. We can then calculate the number
of electrons using the Thomson scattering equations
\citep{billings:66}. If we also know the composition of the CME plasma,
then we can calculate the mass from
\begin{equation} 
M_{CME} = I_{CME}*C_e*C_{plasma} \label{eq:mass}
\end{equation} 
where $C_e$ is obtained from Thomson scattering theory, discussed in the next section, and is the number of electrons per MSB and $C_{plasma}$ the
composition of the CME plasma. Traditionally, a composition of 10\% He
is assumed as a typical value for the corona \citep{1975SoPh...42..163H}. In this case,
$C_{plasma} = 1.97\times 10^{-24}$ g/electron. The actual composition
may vary from event to event. For example, several events contain
filament/prominence structure with likely chromospheric densities and
composition. However, most (if not all) of the cool material is heated
to coronal temperatures quickly \citep{2002SoPh..208..283F}. There is
also very little evidence (e.g. from EUV dimmings) of significant
material ejection from temperatures below a few $10^5$ K
\citep{2008A&A...478..897B}. Coronagraph studies have shown that a
large part of the ejected mass comes from the high corona
\citep[e.g.,][]{1999SoPh..187..427A,2009ApJ...701..283R}. It is
safe to assume that in an average CME, the majority of the plasma
originates in the corona and/or it is fully ionized. But is the value of 10\% He representative
of such plasma? Recent in-situ analysis of ICME composition suggests a
slightly lower contribution of 6\% (Figure 3 in
\citealt{2008ApJ...682.1289R}). To be conservative, we assume the He
composition of CMEs varies between 6\% and 10\% and the expected
uncertainty is then 6\%. So the effect of the composition variability on the mass calculations is of the same order of a few percent as the other factors discussed previously.

\subsubsection{Projection Effects}
The remaining factor in Equation~(\ref{eq:mass}), $C_e$ is also the
one with the largest uncertainties. While the brightness of a single
electron is straightforward to calculate using Thomson scattering
theory, the emission is optically thin. As a result the observed
brightness is the sum of the emission from all electrons along the
line of sight (LOS). Since we do not know \textit{a priori\/} the 3D
electron distribution within the CME, we have to assume a distribution
to derive the number of electrons from the observed excess
brightness. The standard, and simplest, approach is to assume that all
of the emission along a given LOS comes from electrons located on the
Thomson Sphere, which is the location where the LOS is tangent to the
radial from Sun center. For the relatively small elongations covered
by the C3 FOV, and other coronagraphs with outer field limits
$<30R_\sun$, the Thomson Sphere and the sky plane (a sphere of radius
1 AU) are virtually coincident. The concept of the Thomson Sphere was
introduced in \cite{v_h:06} to account for larger distances from the
Sun.

\begin{figure}
\includegraphics[width=5in]{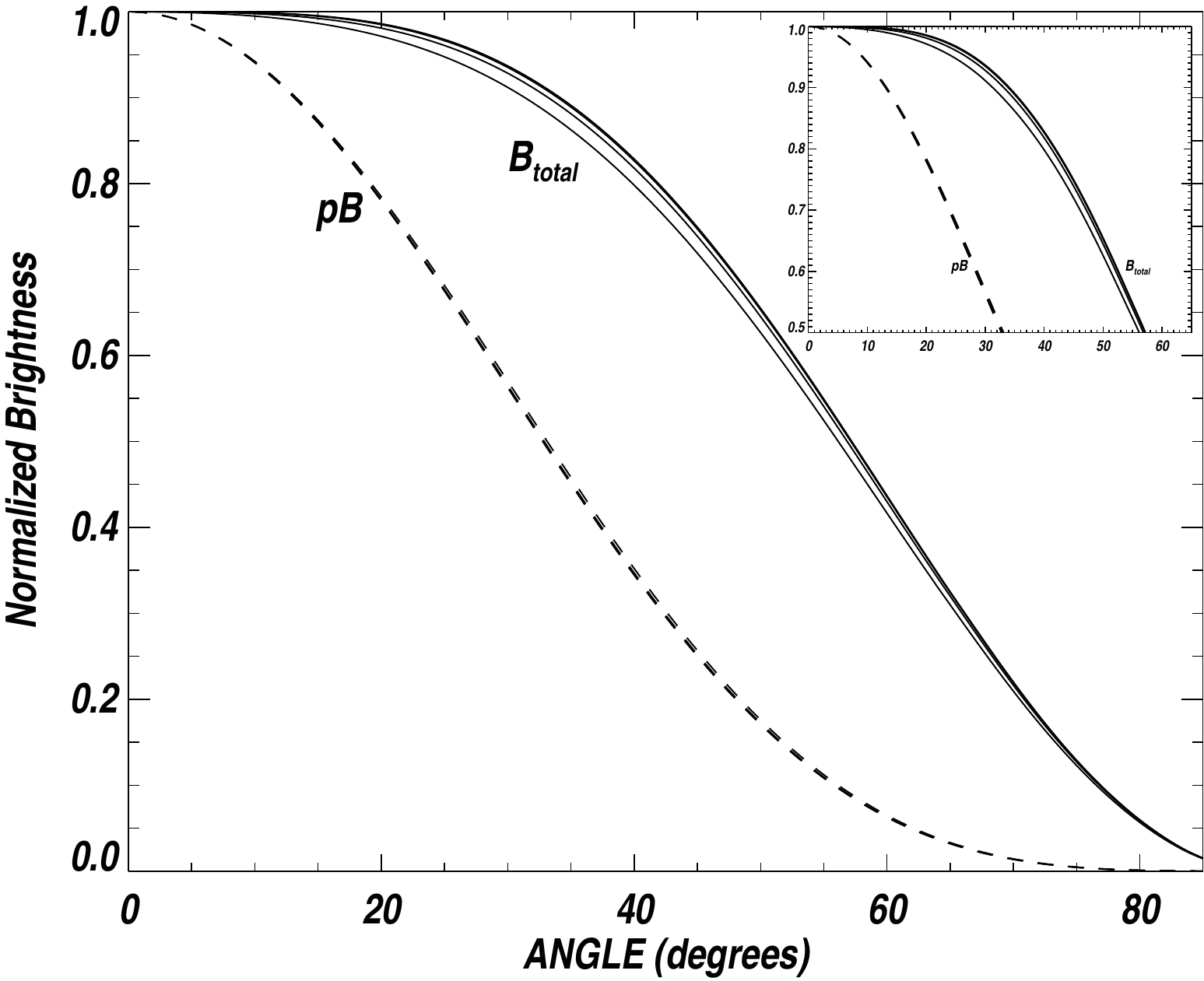}
\caption{Angular dependence of the Thomson scattering brightness of a
  single electron at four impact distances (3, 5, 10, 20 R$_\sun$) for
  total (solid lines) and polarized (dashed lines) brightness. The
  brightness is normalized to the maximum at the sky plane
  (0$^\circ$). The inner curves of total brightness correspond to
  smaller impact radii. The corresponding pB curves are
  indistinguishable from each other. \textsl{Insert:\/} Detail showing
  only brightnesses $>50\%$ of the maximum. See \S~\ref{sec:mass} for
  details.} \label{fig:b_ang}
\end{figure}

In effect, we assume that the three-dimensional distribution of the
CME mass is equivalent to its two-dimensional projection on the image
plane having no longitudinal (along the LOS) extension. Although this
may seem a strong assumption, it has well-understood effects on the
mass calculation. First, it results in a convenient lower bound since
an electron scatters most efficiently on the Thomson Surface, and hence the brightness is maximum at the Thomson surface.  Therefore more electrons are
required at larger angular distances from the Thomson surface to
produce the same brightness (Figure~\ref{fig:b_ang}). Second, the
final mass is underestimated by only $\sim 2x$ as was shown by
calculations \citep{vour:00}, models \citep{2005ApJ...627.1019L} and
more recently by stereoscopic mass measurements
\citep{2009ApJ...698..852C}. 

There is an intuitive way to understand why this simple-minded
assumption provides such robust mass measurements. Let us consider the
angular dependence of the Thomson scattering total brightness
($B_{total}$) in Figure~\ref{fig:b_ang} (see also
\cite{hundhausen_1993}, Appendix A). It is almost constant for the
first 20$^\circ$ from the sky plane, dropping by only 20\% at
$40^\circ$ and finally dipping below 50\% only after 60$^\circ$. Since
the brightness curves are symmetric relative to the sky plane
($0^\circ$), they imply that the total brightness (and hence total
mass) is the same irrespective of how the electrons are distributed
along the LOS within a $40^\circ$ cone. In other words, our assumption
of all electrons on the sky plane results in an exact mass measurement
for any $\leqq 40^\circ$-wide CME propagating along the sky
plane. Since the average CME size is about $40^\circ$
\citep{2004JGRA..10907105Y}, and larger events have a substantial
fraction along the sky plane (with the possible exception of
halo-CMEs), we can see why mass measurements, using the standard
assumptions, represent a reliable estimate of the true CME mass.

\begin{figure}
\includegraphics[width=6in]{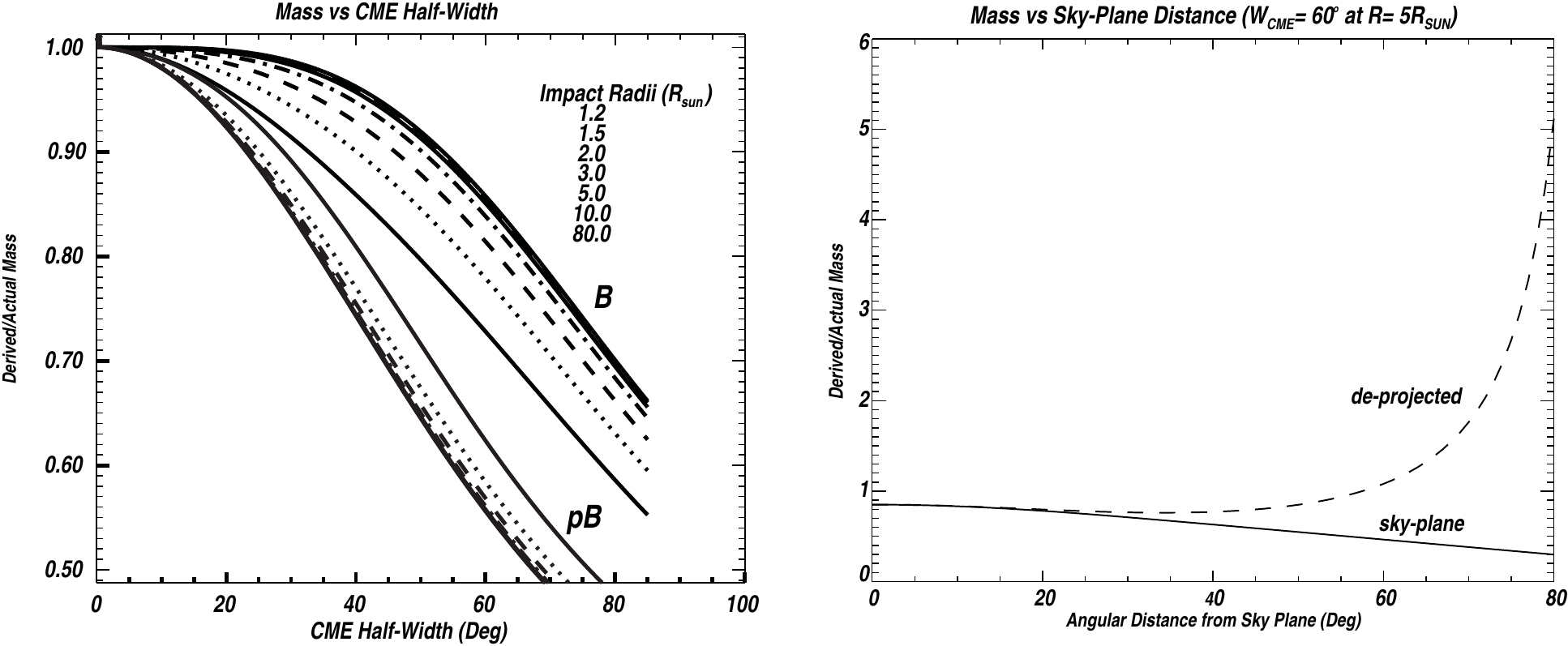}
  \caption{\textsl{Left:\/} Estimation of the error in the CME mass
    as a function of CME half-width. The CME is assumed to be a cone
    propagating along the sky plane. The error for both total and
    polarized brightness measurements and for a number of heliocentric
    distances is shown. \textsl{Right:\/} Estimation of the error in the CME
    mass as a function of CME angular distance from sky
    plane. The estimates are based on a $60^\circ$-wide CME and
    correspond to two cases; the standard assumption of all mass on
    the sky plane (curve labeled 'sky-plane') and taking into account
    the CME angle of propagation (curve labeled 'de-projected'). } \label{fig:error_curves}
\end{figure}

Of course, not all CMEs propagate along the sky plane or have small
widths. For these cases, we can evaluate the reliability of mass estimates using
theoretical 'error' curves. For simplicity, we assume a uniform
electron distribution along the LOS. Then we know the 'actual' CME
mass for CMEs of any width and can calculate the 'observed' mass using the
Thomson scattering equations and the sky-plane assumption.
The ratio between the 'observed' and
'actual' masses is a measure of the expected error in the mass
calculation and is plotted as a function of CME half-width in
the left panel of Figure~\ref{fig:error_curves}. The curves show
calculations at a given projected heliocentric distance (commonly
called impact radius) for both total and polarized brightness. The
inner solid lines correspond to the lowest height (1.2 R$_\sun$). The
curves for heights $>5$R$_\sun$ are virtually indistinguishable from
each other. It is clear that the high angular sensitivity of the polarized
brightness emission leads to significant underestimation of the mass
even for moderately wide CMEs and thus makes it unsuitable for CME
mass analyses. 

In the case of CMEs propagating away from the sky plane,
Figure~\ref{fig:error_curves} (right panel) suggests that the masses
are moderately underestimated reaching a factor of 3x in the extreme
case of a $60^\circ$-wide CME propagating almost along the Sun-Earth
line. It is possible to introduce a first-order correction by assuming
that all the mass lies on a different angle than the sky plane. This
angle should be the CME angle of propagation, if it is known, or if it can be
derived from other observations. An obvious solution is to assume radial
propagation from the source active region
\citep{2003ApJ...590..533R,2005JGRA..11012S01K}. This assumption
improves the accuracy of the mass measurements relative to the
sky-plane assumption out to about $60^\circ$ but diverges rapidly
beyond that angle and overestimates the mass by about 5x at $80^\circ$
(Figure~\ref{fig:error_curves}). Although this is a valid approach for
the analysis of individual events, it is not desirable for a
statistical analysis of large number of events since it results in
both over and underestimation of the masses. Besides, it requires
positional knowledge about the source regions for all the events which
we do not possess. Thus, we retain the 'sky-plane' assumption for our
analysis and noting that the masses of some events (partial or full
halo-CMEs) may be underestimated by as much as a factor of 3.  We note also that the emission from some CMEs is blocked by the occulter reducing the estimate of the total mass of those events.

These considerations affect also the energy calculations. For the
potential energy, the error is the same as for the mass; an
underestimate of a factor of 2. Both mass and speed enter in the
kinetic energy equation and for both quantities the projection effects
are the dominant factors. The most severe speed projection effects
will be associated with CMEs out of the sky plane such as halo and
partial halo events. But we have excluded these events from our
statistics and thus we expect only moderate projection
effects. However, there is no way of knowing which of the thousands of
events in our list are close to the sky plane. To calculate a
conservative error in the speed measurements, we assume a maximum
value of $60^\circ$ of the sky plane which translates to a factor of
$1/\cos{70} = 2$ underestimate of the CME speed. Since both mass and
speed are underestimated, so will be the kinetic energies. Hence, the
kinetic energies can be as much as 8x higher and the potential as much
as 2x. The total mechanical energy will then be as much as 8x. We should therefore consider our mass and energy values as lower limits to the real ones.

To summarize we demonstrated that instrumental effects, the coronal
background and the CME plasma composition are, in general,
insignificant contributors to errors in estimating the CME
mass. Projection effects play the major role but even those can be
reasonably estimated. Our conclusion is that the CME mass and energy
values are lower bounds to the true values and they lead to
underestimations of $\sim2x$ for the mass and potential energy, and
$\sim8x$ for the kinetic energy.

\subsection{LASCO Observations and CME Mass Database}

So far, we have discussed the CME mass measurement procedures and
errors associated with them. The procedures are easily automated and
applied to large numbers of CMEs \citep[e.g.,][]{vour:00}. So we apply
them to all observed CMEs starting from the height-time (HT)
measurements in the LASCO CME database \footnote{\url{http://cdaw.gsfc.nasa.gov/CME\_list}} and we refer to it as the
CDAW CME list hereafter \citep{2004JGRA..10907105Y}. As we describe in
detail below, we have compiled a new database
containing not only mass and energy measurements but also calibrated mass
images for each CME image. We refer to our list as the NRL CME list,
hereafter.  To automate the mass measurements we made some
important decisions at the beginning.

First, it is impractical to outline the CME extent by hand for every
frame for thousands of CMEs, so we measure the total mass
within a sector defined angularly by the reported width and central
position angle (PA) and radially by the occulter edge (lower boundary)
and the reported height of the CME front (outer boundary) plus one $\textrm{R}_\sun$ to ensure that we include the CME front since the
weak brightness of the front may lead to errors in locating it
visually at large heliocentric distances.

Second, we pick a pre-event image automatically based on the time of
the first C2 (or C3) image reported in the CDAW CME list. We search
backwards in time for the first available C2 (or C3) image which is
free of missing blocks along the predefined sector.  If no image is
found up to 6 hour prior, then the event is dropped from the list and
an error log is created. A detailed examination of each CME to determine the best pre-event image (if it exists) is beyond the scope of this paper.

Third, we calculate the mass only for the images used for the
height-time (HT) measurements, even though additional images may
exist. In many cases, not all LASCO data were available at the time of
the height-time (HT) measurements but they may be available
now. Generally speaking, we do not update the mass measurements once
an event has been measured successfully through the above steps. Also
we try to maintain a close correspondence with the HT measurements for
our statistics work, and therefore, we do not want to add mass
measurements where no HT measurements exist.

Four, we want to discriminate between CMEs and solar wind 'blobs'
\citep{1997ApJ...484..472S} as both types of events exist in the CDAW CME
list. It is generally accepted that blobs are the smallest, narrowest
ejections detected in the coronagraphs. However, there is no consensus
on how small can a CME be. Some CMEs can appear narrow because of
projection effects. The properties of events with widths
$<15^\circ-20^\circ$ have been analyzed before with somewhat conflicting
conclusions on whether they form part of the overall CME distribution
\citep{2001ApJ...550.1093G} or not
\citep{2003AdSpR..32.2631Y}. Since it is not clear where the CME width
cuts off and to avoid potential bias we have decided to exclude all
narrow events ($<20^\circ$) from our subsequent analysis. This decision
is further motivated by a sudden upturn on the number of entries in
the CME list starting in 2004. Almost all of the newly added events
are of small width, they are not CMEs and, in our opinion, they are
skewing the overall statistics. Finally, we do not check for accuracy
or re-measure any of the inputs from the CDAW CME list such as speed,
position angle, CME front height, etc. 

\begin{figure}
\includegraphics[width=6in]{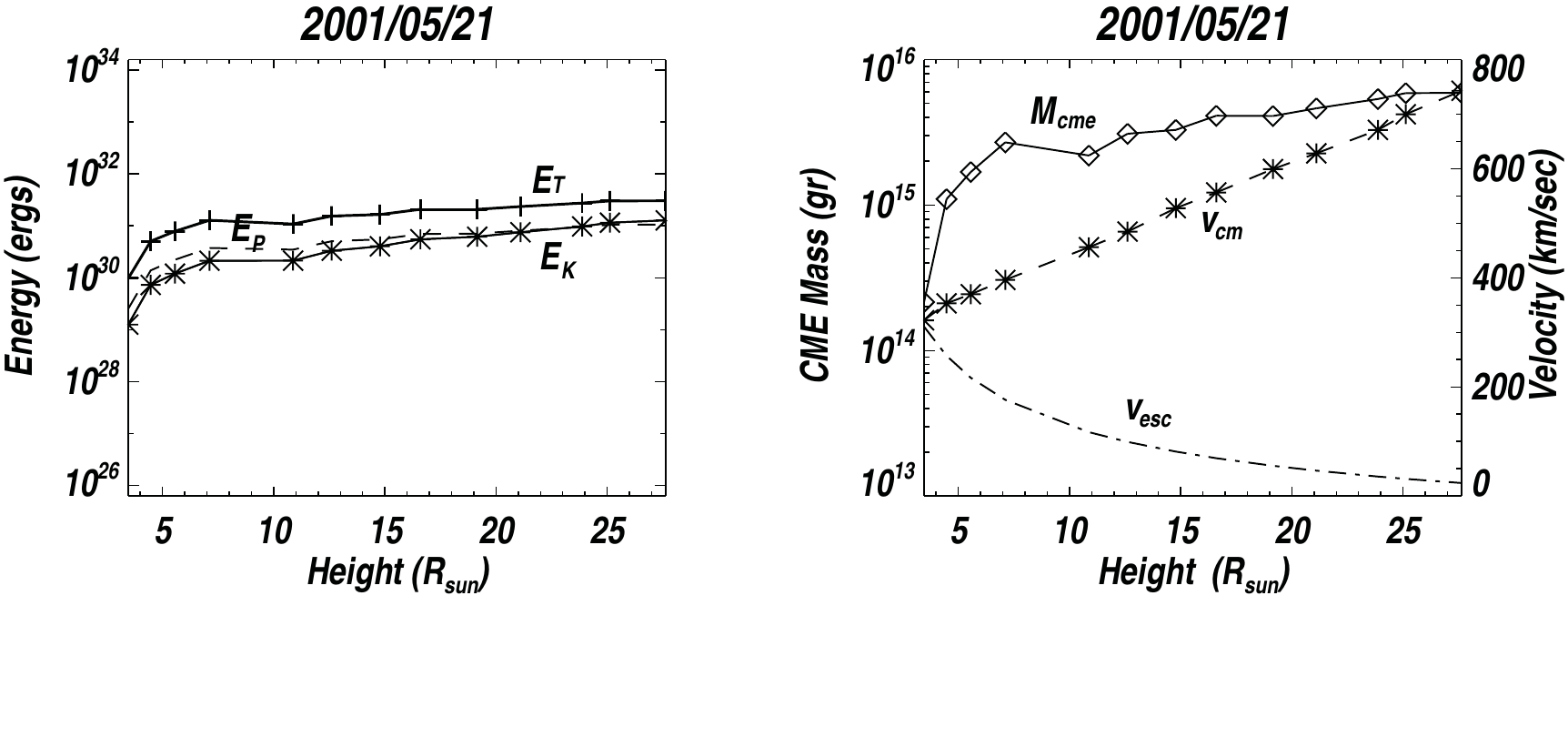}
\caption{\textsl{Left:\/} Height evolution of the CME energy; :
  kinetic ($E_K$), potential ($E_P$), and total mechanical ($E_K +
  E_P$). \textsl{Right:\/} Height evolution of the CME mass
  (M$_{cme}$) and center-of-mass speed (v$_{cm}$). The solar escape
  speed (v$_{esc}$) is also shown. } \label{fig:cme_example}
\end{figure}

The mass CME database is constructed as follows. Our software procedures read in the HT files as they become available
online, retrieve the corresponding C2 or C3 files from the LASCO
database and calculate full resolution, pre-event-subtracted, mass
files (in units of g) which are stored in our mass database at
NRL. For each mass file, we sum the mass within the sector defined
above and calculate the total mass, area, average mass density
(g/pix), center-of-mass height, and potential and kinetic
energies. The details of the calculations of the potential and kinetic energy have
been published in \citet{vour:00} so we will not go into details here. The above
measurements, for all frames of a given event, are stored in a separate
text file. Many additional parameters can now be derived such as speed
and acceleration of the center-of-mass, CME area as a function of
height, etc. To demonstrate the wealth of information contained in our
final CME database, we plot several parameters from a randomly chosen
CME in Figure~\ref{fig:cme_example}.

Here, we focus on the properties of the full CME sample rather than
the evolution of a particular event. So we want to treat each event as
an individual data point and therefore need to extract, for each
event, a representative set of parameters at a single time frame. As
we have shown before, CME mass tends to a constant value above about
10R$_\sun$ \citep{vour:00,2009ApJ...698..852C} and few events show any significant
accumulation out to the edge of the C3 FOV. So it is natural to assume
that a representative point for each event is the time when the CME
achieves its maximum mass. We extract all parameters (speed, mass,
energies, etc) at that time frame and proceed to analyze their
statistics. The final entry for each CME contains the following
data:
\begin{itemize}
\item From the CDAW list: speed, acceleration, position angle, width, front
height. 
\item From the NRL list: mass, number of pixels (summed over to
obtain the mass), average mass per pixel, sector boundaries, kinetic
and potential energies, center-of-mass height and position angle, and
filename.  
\end{itemize}
It is obvious that many statistical studies can be undertaken with so
many available variables, and we looked at several correlations. For the sake of brevity and clarity, we present only the most
important results in the following section.   
   
\section{Statistical Results}
The initial event list contains 14007 CMEs, from 22 January
1996 to 31 July 2009, the last date with HT and mass measurements available
during the writing of this paper. After going through the mass
calculations procedures we end up with a mass sample of 13388 events.
For the analysis here, we need to remove from consideration events
that may bias the analysis. So we remove events with (1) too few
($\leqq 3$) measurements, (2) events with negative mass which may
indicate overlapping events, (3) small width events that may not be
CMEs, and (4) wide events which may have a significant portion of
their structures away from the sky-plane and narrow events. We
therefore consider only events with $120^\circ>$ widths $>
20^\circ$. The remaining sample contains 7668 CMEs, easily the largest
calibrated CME sample ever considered.
\subsection{Duty Cycle Considerations}
For an instrument, the observing duty cycle is defined as the percentage of the total observing time during a certain time period (e.g, day, carrington rotation, etc). For Earth-orbiting
telescopes, which operate over a day-night orbit, the duty cycle is
basically the length of the day orbit. By virtue of its location at
L1, the LASCO instrument is observing the Sun continuously and we only
need to consider interruptions due to instrument or spacecraft
operations. Such corrections are necessary for the evaluation of CME
occurrence rates and are regularly applied
\citep{2000JGR...10518169S,2007AdSpR..40.1042C}. Here, however, we are
interested in a slightly different problem. We do not want to only
know if a CME has occurred but whether it also satisfied the criteria for
inclusion in the database. In the following, we present two other
approaches for calculating the LASCO duty cycle.

\begin{figure}
\includegraphics[width=6in]{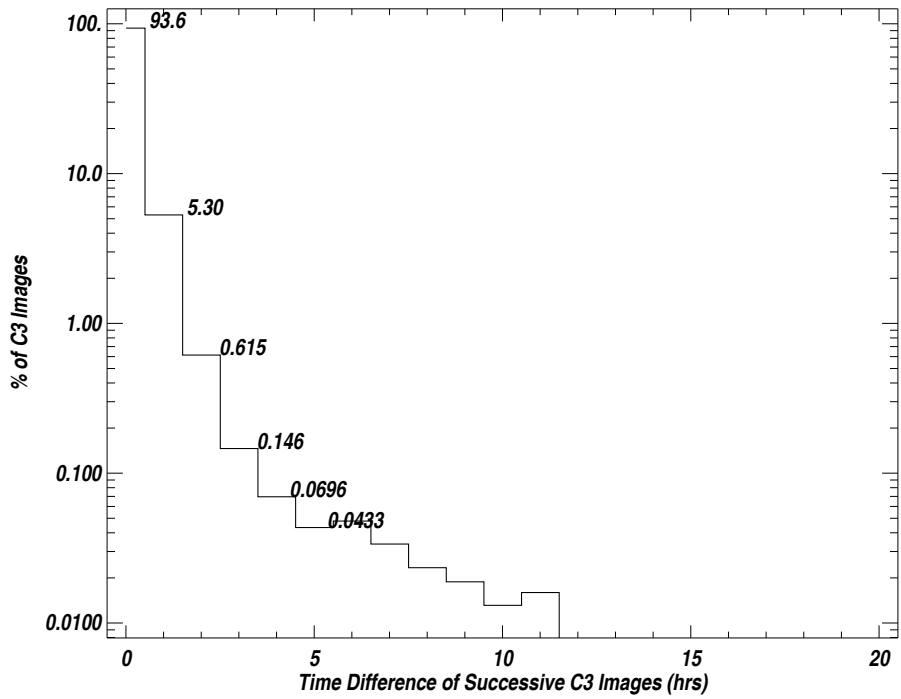}
\caption{Histogram distribution of the C3 image cadence from 1996 to mid-2009. The sample contains all 175,361 synoptic images. The bin size is one hour. The percentage of images in the first five bins is printed on the plot and shows that 98.9\% of all C3 images are taken within two hours of each other.} \label{fig:c3cad}
\end{figure}

For a CME to appear in the catalog, it must have a speed measurement
(an HT plot). For a reliable CME speed, we must have three C3
measurements, at least. Then, the CME will be included, if it has a
speed, 
\begin{equation}
v_{CME} \leq {{\Delta l}\over{n dt}}
\end{equation} \label{c3cad} 
where $\Delta l$ is the distance from the C3 occulter to the edge of the field of view (in km), $n=3$ the number of
images, and $dt$ the time between successive C3 images (cadence, in min). Next,
we calculate $dt$ for all C3 images acquired during the
mission. There are 175,361 such images and we simply take the time difference
between successive images based on the image header. The resulting
histogram, in 1-hour bins, is shown in Figure~\ref{fig:c3cad}. The
y-axis is normalized to the total number of images and represents the
percentage of events in each hourly bin. We see that 93.6\% of C3 images
are taken within one hour from each other. The plot shows that almost all of C3
observations (98.9\%) have a cadence of two hours or less. Setting
$dt=60$ min in Equation~(4), we obtain $v_{CME} = 1824$
km/sec. This result implies that all CMEs
with speeds $<1824$ km/sec will be detected, measured, and will therefore
appear in our statistics. A quick search of the entire CME catalog
shows that only 52 out of 14,409 CMEs (0.4\%) have speeds above 1824
km/sec. While these are important events, eliminating them from the statistics for this paper has an insignificant effect.  Thus, for all practical purposes, 
C3 detects all CMEs with one hour cadence. Therefore,
the overall LASCO duty cycle is 94\%.  

This duty cycle can be used for comparisons with other
instruments (e.g, 66.7\% for \textsl{Solwind\/}) over the length of
the mission. But it does not capture information on finer time
scales. CMEs follow the general activity cycle as a look in
Figure~\ref{fig:cycle} quickly reveals. The number of CMEs per
Carrington rotation is quickly rising during the summer of 1998
(rotations 1938-1940) but there are months with low number of CMEs as
well. Therefore, data gaps may have a varying effect on the statistics
depending on the phase of the cycle. Although we do not pursue this here, future studies may find the duty cycle per Carrington rotation useful information. 

To calculate the duty cycle per Carrington number we proceed as
follows. We showed above that high-speed CMEs will be easily missed
with a few hours observing gap but slow CMEs will not. So we go back
to Equation~(4) and try to estimate the minimum data gap for missing a
CME, irrespective of its speed.

\begin{figure}
\includegraphics[width=6in]{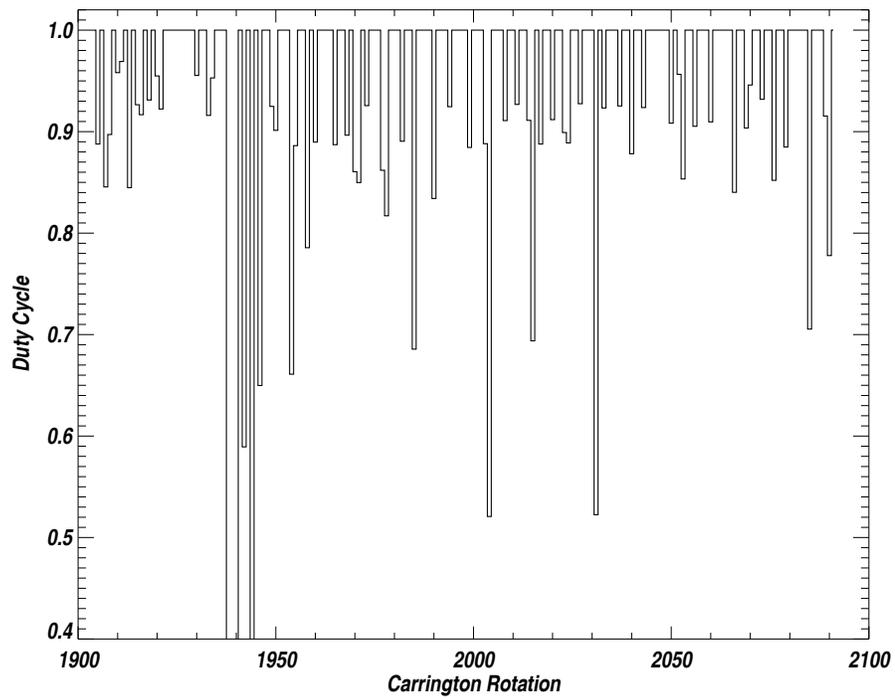}
\caption{The C3 duty cycle per Carrington rotation. The plot is based on the percentage of $> 18$ hrs data gaps. even the slowest CMEs ($\sim100$ km/s) will be missed with such data gap.} \label{fig:carr_duty}
\end{figure}

First, we define a minimum CME speed. We know that the slowest CMEs
are the streamer-blowout types with typical speeds of $\sim 200$
km/sec \citep{vour:02}. The CDAW catalog contains events with speeds
around 100 km/sec but some of them are questionable. There are a
number of outward mass motions that have speeds lower than 100 km/s.
This speed is below the escape speed (216 km/sec), so that outgoing
material slower than this, must be a quasi-equilibrium outflow.  We
believe that these slow events are significantly different from CMEs
which are associated with a magnetic instability and energy
release. In any case, we adopt the rather conservative speed of 100
km/sec which results in $dt=18$ hrs from Equation~(4). Therefore, any
data gap longer that 18 hrs will surely result in failure to detect
all CMEs (for the purposes of inclusion in the
database). Figure~\ref{fig:carr_duty} shows the resulting duty cycle
per Carrington rotation. The long data gaps due to spacecraft
emergencies are immediately obvious but the duty cycle is well within
90\% for the majority of the mission. As a check of our procedure, we calculate the average duty cycle per Carrington rotation for the full time series. We
end up with a duty cycle of 94\% which is the same number calculated above from the time between successive C3 exposures. 


\subsection{Mass and Energy Distributions}\label{sec:histos}

\begin{figure}
\includegraphics[width=6in]{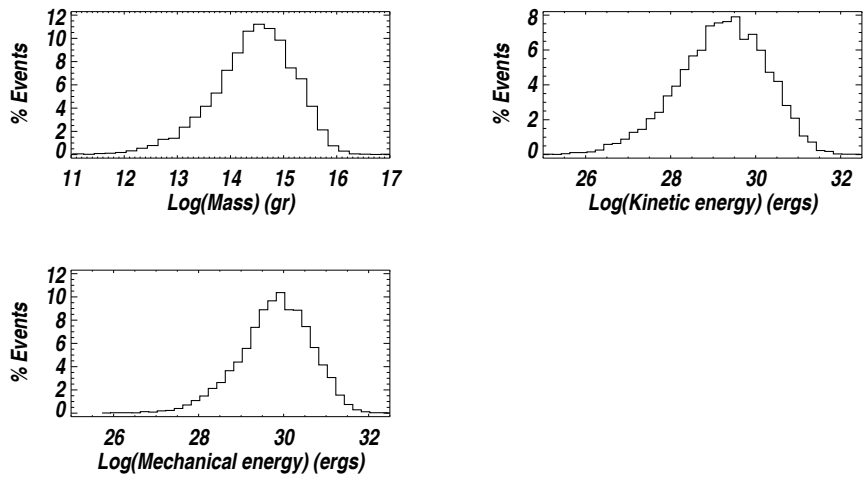}
\caption{Histograms of CME mass distribution (upper left), kinetic
  energy (upper right) and total mechanical energy (bottom left) for
  all 7668 events in our database.} \label{fig:distributions}
\end{figure}

The distributions of mass and energy are shown in
Figure~\ref{fig:distributions}. The bin size is 0.2 dex for all
quantities and the statistics are presented in Table~2. The CME masses extend between $10^{11}$ and $10^{17}$ g
but either extreme is defined by a very small
number of events. 
There are only 7 events with
masses $\geq 5\times10^{16}$ g. At the lower end, there are also only
16 events with masses less than $5\times10^{11}$ g. The low values are
usually caused by overlapping CMEs where the first event creates large
areas of negative brightness against which the second event
propagates. Such cases are impossible to measure reliably in an
automated fashion and are difficult to exclude from the statistics if
their masses are positive. Another contributor to small masses are
events with only one or two frames in C2. We discuss them in
\S~\ref{sec:failed}.  

What type of distribution should we expect for the mass and energy
measurements? Neither the mass nor the energy can be negative and both
have large variances (i.e., the mass values extend over 5 orders of
magnitude). Log-normal distributions arise frequently under
such conditions \citep{limpert_stahel_abbt_2001, crow:88}.  In
addition, log-normal distributions are common in solar physics. They
have been identified in the distribution of magnetic flux in the
photosphere \citep{2005ApJ...619.1160A}, the CME speeds
\citep{2005ApJ...619..599Y}, and CME-associated flares
\citep{2003ICRC....5.2729A}. All three of these parameters are related
to CMEs, either directly or indirectly. It is, therefore, reasonable
to expect the same behavior from CME masses and energies. In that case
the logarithms of both quantities in Figure~\ref{fig:distributions}
should be normally distributed but they are not. The departure from
the normal distribution seems to be caused by an excess of small
values in both mass and energy measurements. The
skewness in the distributions may be caused by either the existence of two populations (a low and
high mass CME population, for example) or by the inclusion of biased samples
 in our statistics. To investigate this, we proceed
as follows.

\begin{figure}
\includegraphics[width=6in]{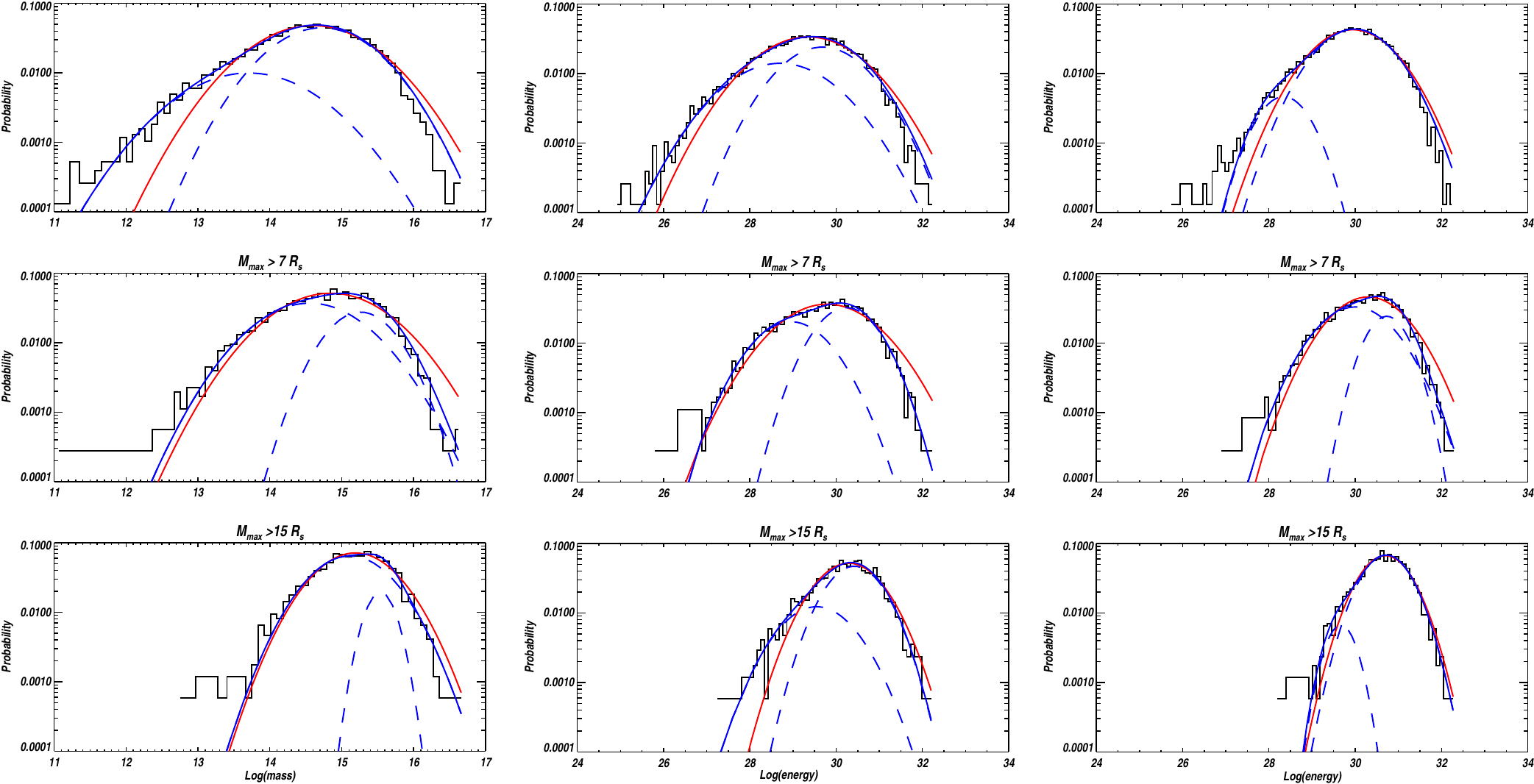}
\caption{Top: Histograms of the logarithms of the distributions of CME
  mass (left), kinetic energy (center) and total mechanical energy
  (right) for all 7668 events. The red lines are single Gaussian
  fits. The blue lines are two-component Gaussian fits and the dotted
  blue lines show the individual Gaussians of the two-component
  fit. Considering events with mass maxima at increasing heights, $7
  R_\sun$ (3579 events, middle), $15R_\sun$ (1704 events, bottom),
  makes the measurements more consistent with a normal distribution of a single population rather than a two-population distribution. See Section~3.2 for
  details.} \label{fig:all_log}
\end{figure}

First, we fit the measurements with a normal curve as shown in
Figure~\ref{fig:all_log}. Mathematically speaking, when the
(normalized) probability function $P$ of the logarithm of a random
variable $x$ is normally distributed with a mean $\mu$, and standard
deviation of $\sigma$ (see also
\citet{2005ApJ...619..599Y,limpert_stahel_abbt_2001})
\begin{equation}
 P = \exp{-({{\log{x} - \mu}\over{\sqrt{2}\sigma}})^2}
\end{equation}
then the variable is log-normally distributed with probability density
function $f(x)$ as
 \begin{equation}
 f(x) = {{1}\over{x\sigma\sqrt{2\pi}}} \exp{-({{\log{x} - \mu}\over{\sqrt{2}\sigma}})^2}
\end{equation}
Here, $\mu$ represents the geometric mean and the one $\sigma$
range is given by $[\mu/\sigma, \mu\cdot\sigma]$ and the $2\sigma$
range by $[\mu/\sigma^2, \mu\cdot\sigma^2]$. Thus we apply a Gaussian fit using equation~(5) to the logarithm of
mass (Figure~\ref{fig:all_log}, top left panel, red line). The data
deviate from the normal fit at both high and low mass values but the excess of
low mass events is especially obvious. 

Then we apply a two-Gaussian fit which
does a much better job in describing the distribution (blue dashed
lines are the individual components, blue thick line is the sum). This
would indicate that we may be dealing with two populations. As we discuss
in detail in \S~\ref{sec:failed}, there are indeed two populations
\textit{when\/} the masses are plotted against their height. 

To check whether the height where the measurement was taken introduces
a bias, we repeat the fitting but exclude all events which reach their
maximum mass at less than 7 R$_\sun$ (3579 events, middle panels) and
at less than 15 R$_\sun$ (1704 events, bottom panels). It is obvious
that the mass measurements approach a log-normal distribution (the
logarithm of mass becomes more normally distributed) as the height of
maximum mass in increased and there is no need for a two-Gaussian
fit. The final parameters of the fit are shown in Table~3. Obviously,
the height where the mass measurement is performed is biasing the
results giving the impression of a two-CME population in
Figure~\ref{fig:all_log}. Therefore, the height, and by direct
implication, the field of view, matters for the proper measurement of
CME properties. This is an important consideration for instrument
builders and for comparisons among different coronagraphs.

We have noted repeatedly in the past that there is considerable
evolution of the CME mass with height
\citep{vour:02,2009ApJ...698..852C}. The mass tends to increase
rapidly between 2-7 R$_\sun$ and reaches a plateau above about 7-10
R$_\sun$ (Figure~\ref{fig:cme_example}). Hence, a maximum mass below
$\sim10$ R$_\sun$ would imply insufficient FOV coverage, overlapping
events or improper event identification (Sect~\ref{sec:failed}). In
all such cases, the measurements should be treated with
caution. Measurements from coronagraphs with limited FOVs (e.g., $\leq
10$ R$_\sun$) are especially affected. We conclude that \textsl{only
  CME measurements to at least 15 R$_\sun$\/ can allow the proper
  measurement of CME properties such as mass and energy}.

We use the same approach for the fitting of the kinetic energy (middle
column, Figure~\ref{fig:all_log}) and total mechanical energy (right
column, Figure~\ref{fig:all_log}). The results of the fit are shown in
Table~3. The behavior is similar but not as striking as in the case of
mass. The log-normal fit is better for the total energy than the
kinetic energy, where an excess of low kinetic energy events is seen
even at 15 R$_\sun$. We do not know the exact reason for this. Since
the kinetic energy is the product of the mass and speed measurements
it should be distributed log-normally. However, we use the final
\textsl{fitted\/} speeds provided in the CDAW catalog and not the
speeds at the height where the mass measurements are taken. In
principle, there should be a small difference, if any, for constant
speed events but a significant one for decelerating or accelerating
events. The excess of small kinetic energies at 15 R$_\sun$ suggests
that their actual speed at 15 R$_\sun$ may be higher than their fitted
constant speed, which is more representative of the speed at 30
R$_\sun$. It is plausible that these events may tend to decelerate in
the C3 FOV. However, a proper answer will require careful speed
measurements for a large number of these events. We leave this exercise for
another paper.

\begin{figure}
\includegraphics[width=6in]{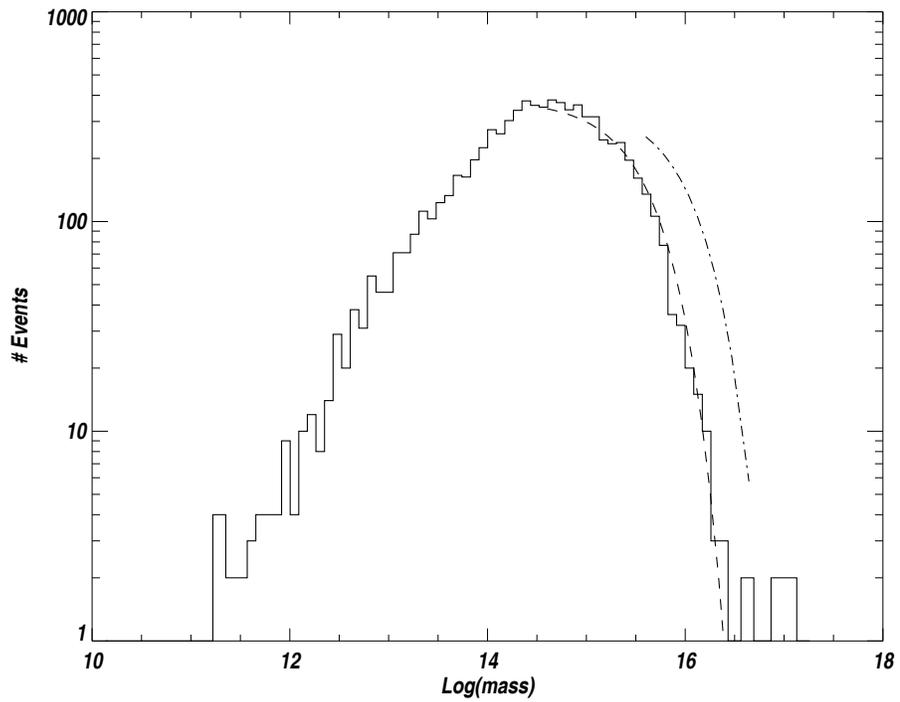}
\caption{Comparison of the exponential CME mass distribution derived by
  \citet{1993SoPh..148..359J} from \textsl{Solwind\/} measurements
  ($4\times10^{16}\geq\mathrm{M}\geq 4\times19^{15}$ g)
  (dash-dotted line) to the LASCO measurements reported here. The dashed
  line is our exponential fit to $4\times10^{16}\geq\mathrm{M}\geq
  5\times10^{14}$ g. } \label{fig:comp_exp}
\end{figure}

We close this section with a comparison of the LASCO mass distribution
to \textsl{Solwind\/} distribution \citep{1993SoPh..148..359J}. This
is the only other published mass distribution we are aware
of. \citet{1993SoPh..148..359J} used $\sim1000$ CMEs to construct a
mass distribution between $10^{15} - 5\times10^{16}$ g. Their
distribution for M$\geq 4\times10^{15}$ g was described well by the
exponential law $N_{CME} = 370 e^{-9.43\times10^{-17}M}$. We compare
their fit (dash-dotted line) to the LASCO mass distribution in
Figure~\ref{fig:comp_exp}. Clearly, it is biased towards higher masses
by about half a decade. This is expected, given the lower sensitivity
of \textsl{Solwind\/} relative to the LASCO coronagraphs. The dashed
line shows a better exponential fit between $4\times10^{14} -
5\times10^{16}$ g. The lower cutoff is taken at the mass value where
the distribution dips to lower values and is one order of magnitude
smaller than the \citet{1993SoPh..148..359J} cutoff. The new law is
$N_{CME} = 370.1 e^{-2.3\times10^{-16}M}$ which is a factor of 3.9 below the \textsl{Solwind\/} fit and leads to correspondingly lower mass
fluxes (Table~\ref{tbl:properties}). \citet{1993SoPh..148..359J}
justified the exponential fit and the existence of a detection
threshold (assumed at $\sim4\times10^{15}$ g) on the grounds that the
downturn in the distribution towards the lower values was due to
instrument sensitivity. As we demonstrate in this paper, the LASCO
coronagraphs are perfectly capable of detecting events carrying as
little as a few $\times10^{12}$ g so the observed downturn at around
$4\times10^{14}$ g is not due to instrumental effects but is
intrinsic to the mechanism of the generation of CMEs. Although the
high mass values are described well by an exponential, we see no
reason to use anything other than a log-normal fit.
 
The difference in the mass distributions between LASCO and
\textit{Solwind\/} could be due to a systematic error in the mass
estimates or it could reflect a real change in the average CME mass
between the two solar cycles.  The \textit{Solwind\/} measurements
incorporated CMEs of all widths, including halos, which might affect
the histogram.  Also \textit{Solwind\/} CME masses were computed using
a proxy based on the width of the CME and the apparent surface
brightness.  The method was described in \citet{russ:85}.  Briefly,
the mass of a small number of CMEs (15 events) was determined with the
same method of preevent subtraction as the LASCO masses.  Then the
mass of each CME was divided by its width, which resulted in a mass
per degree.  An average mass per degree was computed for the three
apparent brightness groups (Faint, Average, Bright).  Those averaged
mass per degree values were used to estimate the mass for all the
Solwind events.  The \textsl{Solwind\/} masses are tied to a set of
events that were calibrated in the normal sense, and are consistent
with the masses calculated with the \textit{SMM} and \textit{Skylab\/}
coronagraphs. It seems very unlikely that the event selection or a
systematic offset can explain the difference in the mass curves of
Figure~\ref{fig:comp_exp} while maintaining the same exponential
behavior. We have to conclude that the CMEs observed by LASCO,
while having the same shape to the distribution function are in fact
\textsl{less massive than the CMEs from 1971-1990\/}. The downward
trend in CME properties is reflected in the transition to cycle 24 as
we will see in \S~3.5.

\subsection{The 'Average' CME}
\begin{figure}
\includegraphics[width=7in]{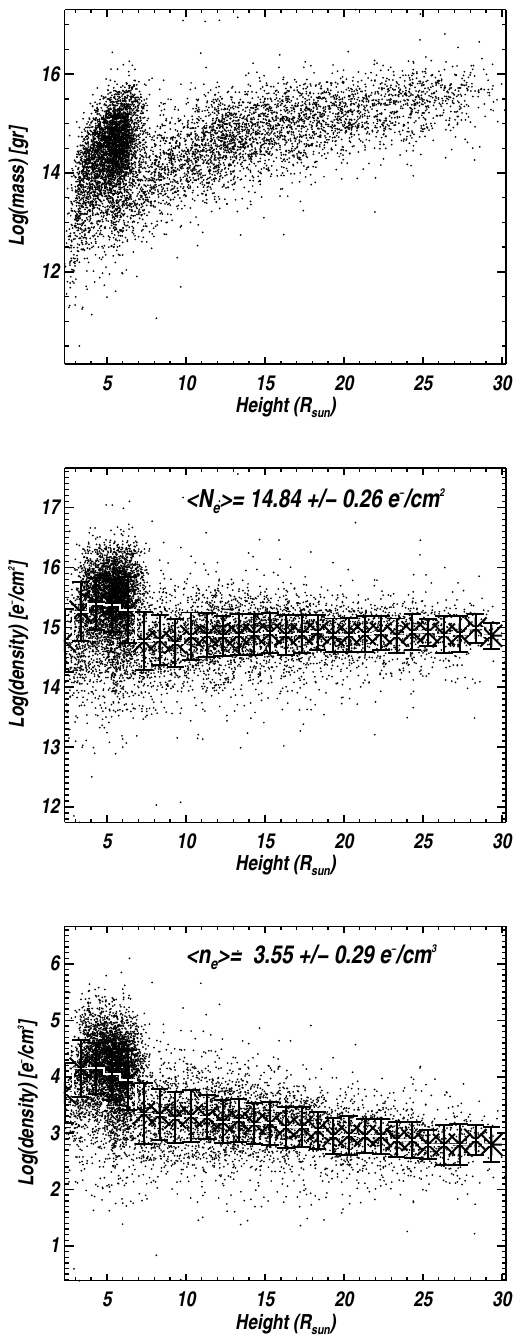}
\caption{\textsl{Top:\/} Scatterplot of the logarithm of maximum CME mass
  versus the height where it was measured. Two populations are
  present: CMEs reaching maximum mass $<7$ R$_\sun$ and CME with
  maximum mass $>7$ R$_\sun$. The former are discussed in
  \S~\ref{sec:failed}. \textsl{Middle:\/} Scatterplot of the logarithm
  of CME surface density (e/cm$^2$) versus height. The two populations
  are clearer here. The CME density is constant above
  $\sim10$R$_\sun$.  A histogram with 1 R$_\sun$ bins is calculated
  and the average density (asterisks) and standard deviation (error
  bars) in each bin are overploted. The small spread of the CME density values
  above $\sim10$R$_\sun$ is quite remarkable. Its average value is
  shown on the plot. The height spread is mostly due to the noise and
  flatness of the mass measurements at those heights which tend to
  shift around the height of the maximum mass. \textsl{Bottom:\/} CME
  volume density versus height. The density is derived assuming a LOS
  depth equal to the projected width. The slow decline of the density
  towards the outer FOV is consistent with the reduced SNR at these
  heights. There is no evidence of material pileup.
}\label{fig:scatterplots}
\end{figure}

We note an interesting behavior in the height distribution of the
maximum mass (Figure~\ref{fig:scatterplots}, top panel). The data
points seem to separate into two populations. There is a 'low corona'
population located mostly in the C2 FOV, at heights below 7R$_\sun$
with higher masses and a population spread throughout the rest of the
C3 FOV showing a slightly increasing mass with height. The
latter is representative of the mass evolution of individual CMEs
(Figure~\ref{fig:example}). The 'low corona' component is, however,
unexpected. Part of the mass increase with height is due to the
increase of the CME area as it expands in the outer corona (see previous section). To account
for that, we divide the masses with the number of pixels used to
measure them and then transform the area from pixels to physical units
(cm$^2$). Thus, we obtain the CME column mass density (g/cm$^2$) and
column electron density (e/cm$^2$). We plot the latter in the middle
panel of Figure~\ref{fig:scatterplots}. As we suspected, the mass
increase was due to the increasing, i.e. more of the event becomes visible in the coronagraph field of view. CME area. The CME density
curve is flat above about 7 R$_\sun$ with a scatter of only a factor
of 10. Given the large number of event sizes, orientations and
projections, the degree of flatness seems remarkable. The height
spread can be easily understood when we consider that the CME mass
tends to a constant value above about 10 R$_\sun$
(Figure~\ref{fig:cme_example} and
\citealt{vour:00,2009ApJ...698..852C}). This implies that the rapid
expansion and post-CME flows have largely ceased and we call
the event 'mature' at this point. However, the mass measurements are
not absolutely flat but have some fluctuations. Since we pick the
maximum mass automatically, this 'noise' will naturally result in a
height spread. A more sophisticated selection criterion, such as
fitting the individual event mass curves and picking the 1/e values
(as in \citealt{2009ApJ...698..852C}), should result in a much smaller
spread.  

Next we bin the data above 10 R$_\sun$ in 1 R$_\sun$-wide bins and
derive the average and standard deviation of the quantities in each bin. Fitting
these points with a straight line results in the average density
\begin{displaymath}
\langle N_e \rangle = 10^{14.84 \pm 0.26}\  \textrm{e}^- \textrm{cm}^{-2}
\end{displaymath}
or in terms of mass density, $\rho$ 
\begin{displaymath}
\langle \rho \rangle = 10^{12.83 \pm 0.25}\  \textrm{g/R}_\sun^2
\end{displaymath}
We see that the spread in the densities is only a factor of about 3 which is much less than the four order of magnitude spread in the measured CME
masses.  It clear that, once the event has 'matured', the average CME
density is rather constant and the observed variation in masses and
consequently shapes and morphologies is mostly the result of
projection effects rather than properties of the source region.

As an extension of the above discussion, we can estimate
the volumetric CME electron density assuming that the LOS depth is equal to the
projected angular CME width. For a sector with radial extent between
radii, $R_1$, and $R_2$ and angular extent between position angles,
$\theta_1$, and $\theta_2$, or width, $\Delta\theta = \theta_1
-\theta_2$, the CME volume, $V_{CME}$ is given by 
\begin{equation}
v_{CME} = {{2}\over{3}} (R_2^3 -
R_1^3)\Delta\theta\sin{{(\theta_2-\theta_1)}\over{2}}
\sin{{(\theta_2+\theta_1)}\over{2}} 
\end{equation}
The result in shown in the bottom plot of
Figure~\ref{fig:scatterplots}. The volume density plot is similar to
the other two except for a slight
negative slope towards larger heights. This is expected on Thomson
scattering considerations. As the CME expands, it becomes fainter due to the $R^{-2}$ dependence of the scattering, especially for the parts away from the sky plane. When their brightness reaches background levels, it stops contributing to the total CME brightness. The very
shallow drop, however, even at 30 R$_\sun$ is a testament to the high
sensitivity of the C3 coronagraph. The average CME electron density is then 
\begin{displaymath}
\langle n_e \rangle = 10^{3.55 \pm 0.29}\  \textrm{e}^- \textrm{cm}^{-3}
\end{displaymath}
Again, there spread in the volume density is only a factor of
3.8. This result, based on statistics of several 1000s of events,
suggests, rather strongly in our opinion, that there is very small
variation in the mass densities of the ejected plasma and that the
plasma must come from higher altitudes where there is less density
variation compared to the low corona. For example, if we assume that
CMEs expand adiabatically in the corona, their density should drop as
$R^{-3}$. Since we find that the events reach a more or less constant
density at $10-15$ R$_\sun$, their original density, on average, on
the surface ($\sim 1$ R$_\sun$) should be $\sim 3.5-12\times10^6
\textrm{cm}^{-3}$. However, these values are 2 to 3 orders of
magnitude less than the densities routinely found in loops from EUV
and SXR diagnostics. Therefore, it seems unlikely that the majority of
the plasma seen in white light CMEs at 15 R$_\sun$ originates from the
low corona, if the adiabatic expansion assumption is valid. Moreover,
the EUV images of the low corona show that densities can vary
significantly (much more that an order of magnitude) from active
region to active region and between active and quiet sun. This is not
what our measurements show suggesting that plasmas at larger scales
are involved.

The spread in the CME densities could be caused by departures of the
CME shape from the spherical symmetry implied by the assumption of a
LOS depth equal to the projected width. We can associate the density
spread to a variation of the CME aspect ratio, $K$, defined as the
ratio between its projected width over its LOS depth. In that case,
the variation in $K$ is proportional to the variation in density and
therefore $K \propto 3.8/2 = 1.9$, in the outer corona. Interestingly, this value is
consistent with forward modeling ratios in \citet{2009SoPh..256..111T}
($K = \kappa/\sin\alpha$ in their notation) and flux-rope fitting by
\citet{krall_st.cyr_2006} ($K\approx (1-\epsilon^2)^{1/2}\Lambda$ in
their notation).


\subsection{'Pseudo-CMEs'} \label{sec:failed}

We now return to Figure~\ref{fig:scatterplots}. Why are there events
that do not show a continuous mass increase as they expand in the
outer corona but rather peak within $7\textrm{R}_\sun$?  We can think of several possibilities for such measurements: (i) unavailability of
observations at higher heights (i.e., instrumental problems,
spacecraft operations, special observing programs). (ii) Multiple
event overlap which makes it difficult to follow the original event. This
may be common during the solar activity peak. (iii) Extremely fast
events which allow only a handful of observations in the LASCO
coronagraph FOV. (iv) CMEs producing large amounts of prompt particles
which in turn create a cosmic ray 'storm' on the detectors and hinder
measurements, and finally (v) normal events that start bright in the
C2 images but diffuse and later disappear in the C3 FOV for various reasons such as (v1) material draining back. (v2) They are located off-limb, so that the brightness falls off faster than the background. (v3) They expand super-radially, thereby reducing the volume density, and (v4) part of the emission originates not from Thomson scattering but from other processes, most likely H$\alpha$ emission. 

\begin{figure}
\includegraphics[width=6in]{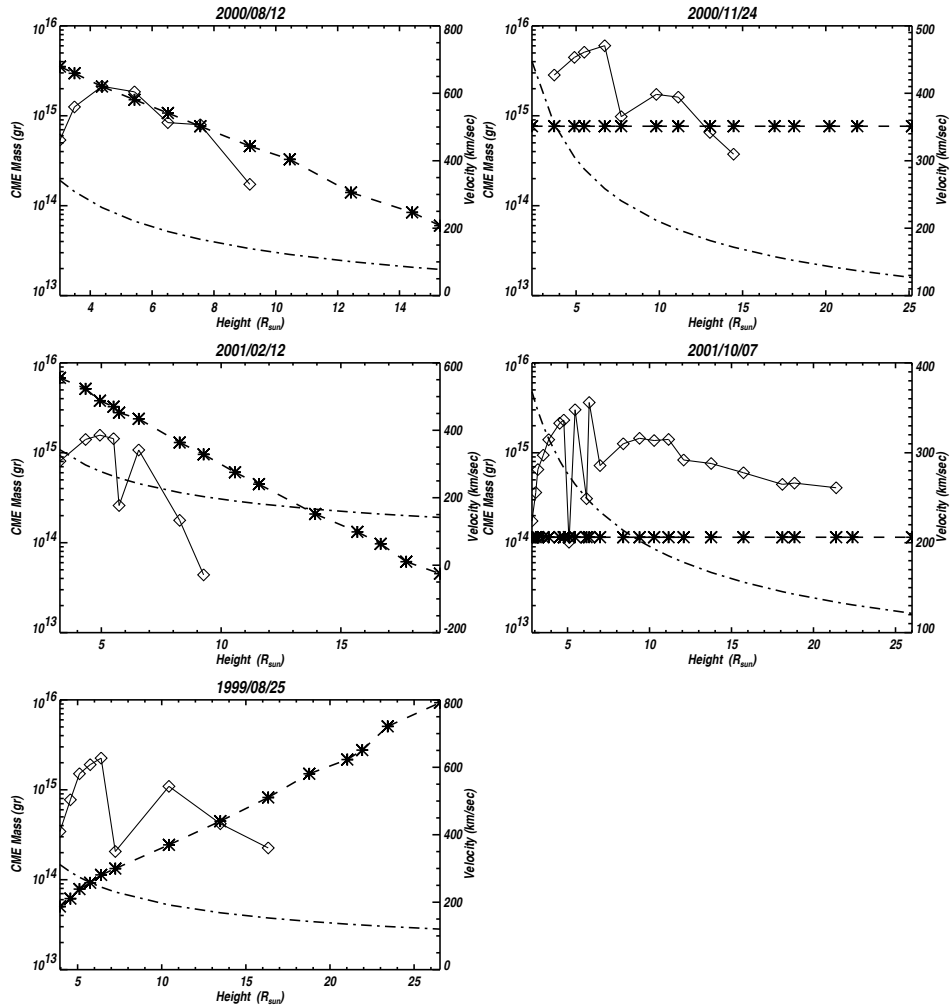}
\caption{Mass and speed evolution of events selected from the
  'pseudo'-CME population. The mass declines with height after
  reaching a peak in the inner corona but the events have speeds above
  the escape speed. Mass (solid line + diamonds), center-of-mass speed
  (dash + asterisks), escape speed (dash-dot). The difference in the
  number of mass and speed data points is due to the plotting of all
  speed but only positive mass measurements. }\label{fig:failed2}
\end{figure}

The large
number of these events (54\% of the total sample) makes them unlikely
to be due solely to erroneous measurements or missing
observations. Besides, we have excluded from consideration all events
with $\leqq 3$ HT points and all halos and partial halos which tend to
be the fastest CMEs and most likely to be associated with particle storms. Therefore, some of these events exhibit a real decline of
mass with height, either because of overlap with other CMEs or because of reasons (v1) to (v4) outlined above. Indeed, this is what we find when we look at some
individual events in more detail (Figure~\ref{fig:failed2}). After an
initial peak in the C2 FOV, the mass seems to decline rather rapidly
until the event reaches background brightness and disappears within
the FOV. Since we plot only the positive mass measurements, these
points end well before the available height-time measurements
(asterisks). An inspection of the mass images reveals quickly whether this is due to an event expanding into another CME. In that case, the preevent image may contain part of the preceding CME and will therefore have negative values in those locations which will result in negative measurements (e.g., 2000/08/12 and 2000/11/24 events). For the other two events, however, we find nothing unusual in the images. The CME simply becomes fainter as it expands in the C3 FOV. This behavior is different from usual CMEs (e.g.,
Figure~\ref{fig:cme_example} and \citealt{vour:00}) and thus we suspect that the nature of these events may be different. They may not be magnetically driven, for example. However, we cannot rule out the contribution from H$\alpha$ within the C2 FOV since the C2 bandpass includes the H$\alpha$ line. Many CMEs clearly contain filamentary material which is likely H$\alpha$ when it is quite luminous (usually in small knots) in low heights. The material ionizes quickly and there are rarely any traces in the C3 FOV. Since our measurements are performed automatically, and many CMEs contain filamentary material  it is reasonable to assume that for at least some CMEs the mass peak in low heights must be due to contributions from H$\alpha$. 

\begin{figure}
\includegraphics[width=6in]{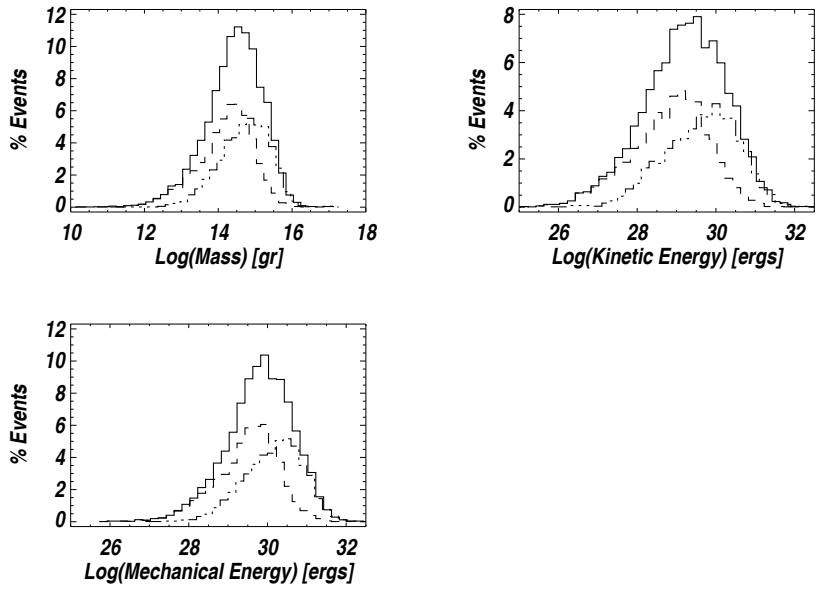}
\caption{Same as Figure~\ref{fig:distributions}. Also shown are the
  histograms for events reaching maximum mass $<7$R$_\sun$ (dashed
  lines) and events reaching maximum mass $\geqq7$R$_\sun$
  (dash-double dot). }\label{fig:failed}
\end{figure}

Much more work is required to properly discriminate among all these possibilities and determine whether the number of events that are truly different than normal CMEs is significant. We, tentatively, refer to all events with mass peaks below 7 R$_\sun$ as 'pseudo-CMEs' to differentiate them from the rest of the sample since they do have very different properties. In Figure~\ref{fig:failed}, we compare the mass and energy
distributions of the two populations to the overall distributions of
the sample. It is clear that the 'pseudo-CMEs' account for the low end
tails of the distribution and have smaller masses and energies than
the events above $10 \textrm{R}_\sun$. A detailed analysis
of these events requires a deeper 'cleaning' of the database to remove
instrumental and other effects, as we mentioned above, and further
examination of their evolution. Such analysis will detract from the
focus of the current paper and will be presented in a forthcoming
publication. The current discussion aims to make the reader aware that
such events exist and that they may not be CMEs or, at least, not CMEs in the way we are accustomed to think of them.

\subsection{Solar Cycle Effects } \label{sec:cycle} 

\begin{figure}
\includegraphics[width=6in]{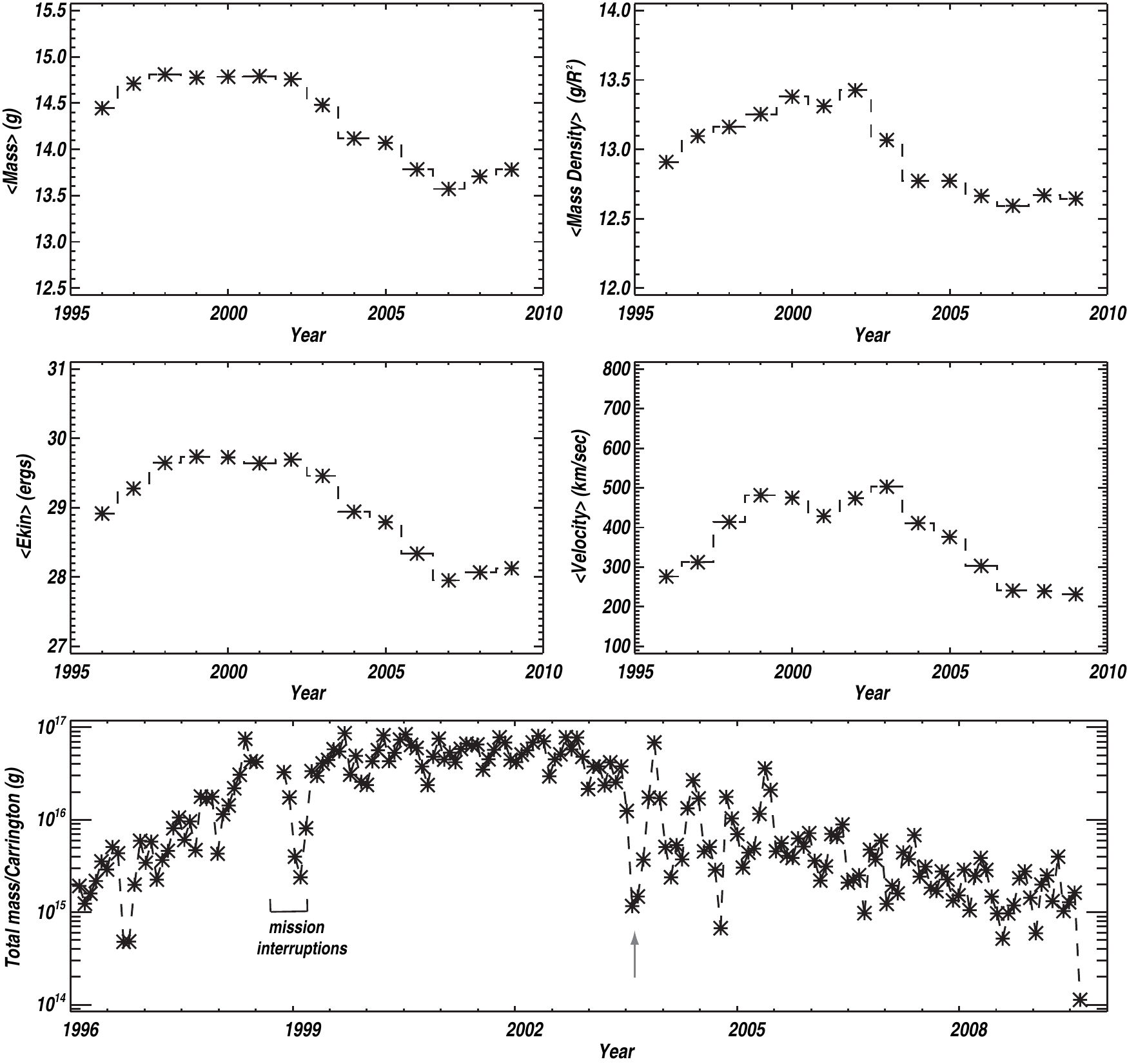}
\caption{The solar cycle dependence of the CME mass and kinetic
  energy. \textsl{Top left:\/} Log CME mass.  \textsl{Top right:\/} Log CME
  mass density in g/R$_{\sun}^2$. \textsl{Middle left:\/} Log CME kinetic
  energy. \textsl{Middle right:\/} CME speed. All four plots show yearly
  averages. \textsl{Bottom panel:\/} Total CME mass per Carrington
  rotation. The data gaps in 1998 and the drop in 1999 are due to
  spacecraft emergencies. The pronounced change in the mass
  variability (arrow) after the middle of 2003 is of solar origin but not fully understood yet.}
\label{fig:cycle} 
\end{figure}

There have been numerous studies of CME properties in the past but
they never extended over a full solar cycle. The almost uninterrupted
LASCO observations for 13 years (so far) provide us with an
unprecedented opportunity to study the variation of the CME properties
through more than a cycle. CME properties such as speeds, widths,
position angles and rates over various lengths of the recent solar
cycle 23 have been presented before \citep{2004AdSpR..34..391G,
  2004JGRA..10907105Y, 2007AdSpR..40.1042C, 2009EM&P..104..295G}. Here
we concentrate on the mass and energy properties, expanding on our
earlier report \citep{vour:02}. There has been an indication for a
solar cycle dependence on the CME mass. \citep{2001ApJ...549.1175M}
derived a factor of 4 increase in the CME mass between the minimum and
maximum of cycle 23 but these results are based on partial cycle
coverage.  We plot the yearly average of CME mass as a function of
time in the upper left panel of Figure~\ref{fig:cycle}. It is rather
clear that the mass depends on the cycle although it remained
relatively flat between 1997-2002. The average begins to decline in
step-wise fashion in 2003 until 2007 when we start seeing an increase
in the average mass. The current 2009 average ($6\times10^{13}$ g) is
still much lower (\textsl{by $4.6\times$\/}) than the average in the previous
minimum ($2.8\times10^{14}$ g) in 1996.

This behavior becomes more visible if we consider the average
mass density (upper right panel, Figure~\ref{fig:cycle}). There we see
a continuous increase in the average CME density as the activity
increases. Even the characteristic double-peaked shape of the sunspot
number is reproduced in the plot. Starting in 2003, however, there is
a sharp decline leading to a plateau in 2006. There is only a very
small hint of increasing density in the 2009 data. 

The decline does not correlate with the
measured speed (middle right panel, Figure~\ref{fig:cycle}) which
shows a good correlation with sunspot numbers. However, the average
speed in 2009 (231 km/sec) is also lower than the average 1996 speed
(276 km/sec) but the difference is much less pronounced than in the
CME mass and density. 

The CME kinetic energy, being the product of mass and the square of
the speed, reflects the variation of both quantities. However, the
mass varies by about a factor of 10 throughout the cycle and the speed
varies only by a factor of $\sim2$. So the kinetic energy is more
influenced by the mass variations. There is a decline in CME energies
in 2003 and a minimum in 2007. The current average kinetic energy
($1.3\times10^{28}$ ergs) is \textsl{$6.4\times$ smaller\/} than the
average CME kinetic energy in 1996 ($8.3\times10^{28}$ ergs).

\subsubsection{The Sharp Decline in CME Mass in mid-2003}

To the best of our knowledge these behaviors have not been discussed before. To
see whether this is an effect from the yearly averaging and to locate
the time of the sharp decline in CME mass more precisely, we compute
the total CME mass per Carrington rotation in the bottom panel of
Figure~\ref{fig:cycle}. The plot corroborates our yearly statistics
and shows some very interesting trends. The mass increases with solar
activity in an intermittent fashion of 2-3 months of high mass
followed by periods of lower output (see also Figure~3 in
\cite{vour:02}). The peak is reached in April-May 1998 and the monthly
mass remains more or less constant thereafter. The drop seen in the
late 1998 - early 1999 is due to the temporary loss of the SOHO
spacecraft and other spacecraft emergencies until synoptic
observations were reliably reestablished in March 1999. The monthly mass
starts to decline in 2003 but then it suddenly drops by a factor of 10
between Carrington rotations 2004 and 2005 (marked by an arrow in
Figure~\ref{fig:cycle}).  This corresponds to the months of June and
July of 2003. 

We do not yet have a fully satisfying explanation for the sudden drop
in 2003. It is not seen in the projected CME widths or the heights but
there is a small hint of it in the ROI area. Our first suspect was the
magnetic field polarity reversal. But that occurred in 2000-2001 and
affected only the position angle of the events, apparently
\citep{2003ApJ...598L..63G}. The drop is very apparent in the mass
density plot and less apparent on the energy plot. This is due to the
different speed behavior. The CME speeds follow closely the two-peaked
evolution in the sunspot numbers (Figure~\ref{fig:cycle}). This
discrepancy suggests that the drop is more of a coronal density rather
than a magnetic effect. Our ongoing analysis of the streamer
brightness throughout the solar cycle shows the same behavior as the
CME masses and corroborates our suggestion
\citep{2009AGUFMSH13C..04H}. On the other hand, the steepness of the
drop may be exaggerated by observing interruptions due to spacecraft
antenna problems during that period. SOHO started its 3-month
$180^\circ$ rolls at the end of the summer 2003. It may also be
normal. The activity also peaked relatively quickly in 1998 during the
rising phase of cycle 23. we do not have CME measurements over a
sufficient number of solar cycles to make a definitive statement. We
will continue to investigate this issue and will report our findings
in a future publication.

\subsubsection{Six-month variability in CME mass}
After 2003, the CME mass shows large fluctuations over the scale of a
few months until about 2006. The variability seems to be related to
the appearance of large sunspot groups (e.g. ARs 10486-488 in the
October-November 2003 period) but it also seems to be periodic. To
look for periodicities in the mass data, we use the Lomb Normalized
Periodogram (LNP) method which is robust against data gaps and readily
available in IDL. The procedure has been developed as an alternative
to Fourier transform analysis for working with unevenly spaced data
and time series with data gaps
\citep{1976Ap&SS..39..447L,1982ApJ...263..835S} and
is therefore appropriate for our data sample. The algorithm accepts as
input the amplitude of a signal versus time, computes the periodogram
and tests the hypothesis that a given frequency peak represents a
significant periodic signal against the white noise hypothesis. The
program returns the power of the signal as a function of frequency and
the probability for each peak of arising from random noise. A low
probability value denotes a significant periodic signal. We use the
method as implemented in the IDL LNP\_TEST routine.

\begin{figure}
\includegraphics[width=6in]{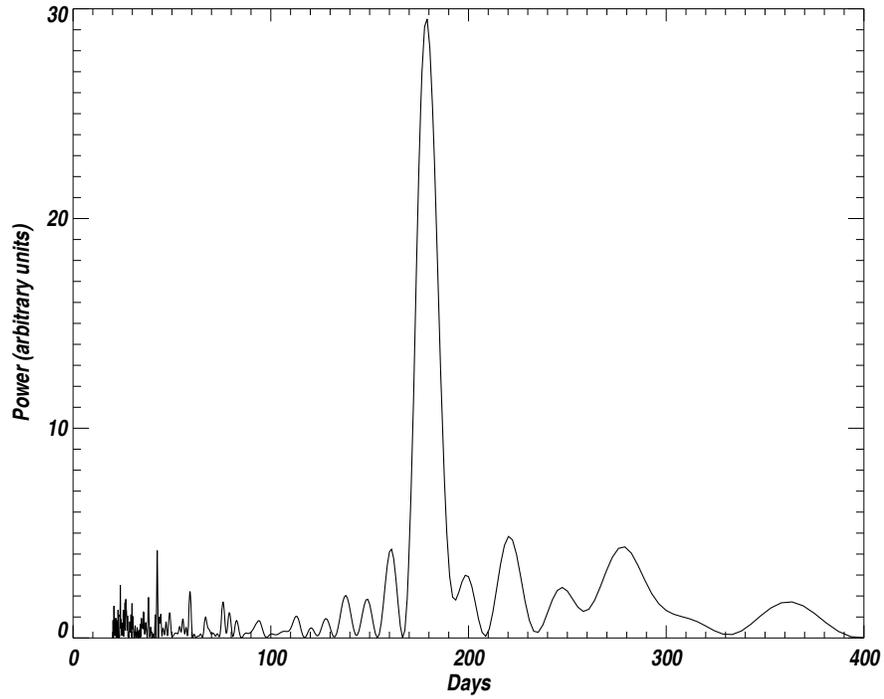}
\caption{Lomb-Normalized periodogram for CME mass measurements in 2003
  - 2009. The x-axis is labeled in days and the y-axis is the power of
  the peak. The measurements are bined in 10-day bins. There is a very significant peak at 179.1 days. The
  probability of being a random peak is
  $3.7\times10^{-11}$.}\label{fig:lnp}
\end{figure}

\begin{figure}
\includegraphics[width=6in]{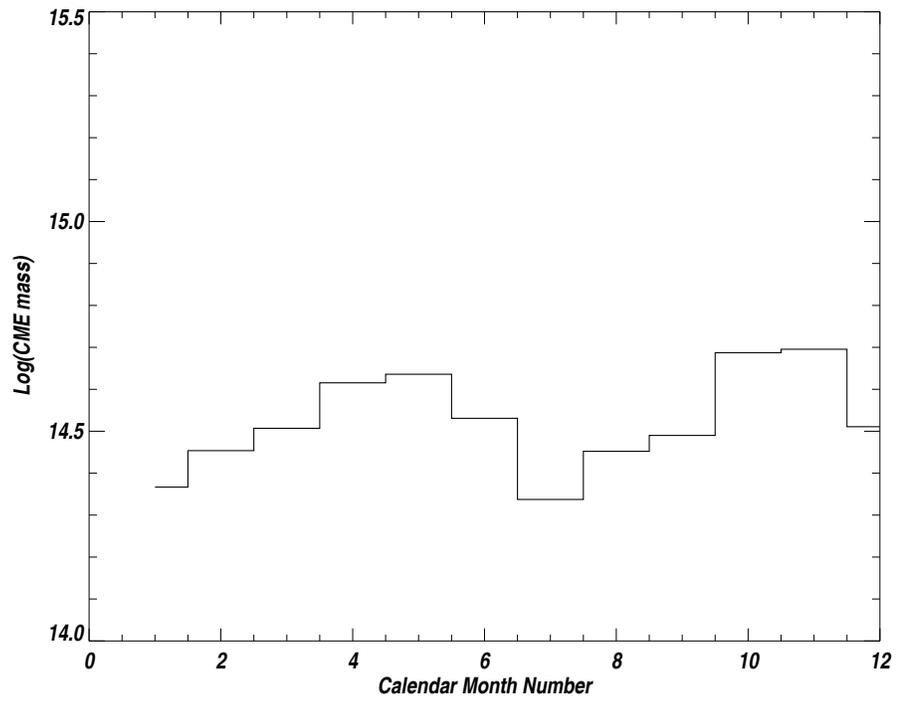} 
\caption{ The logarithm of the average CME mass per calendar month is
  plotted for the full CME database.  The graph suggests that the
  6-month variability shown in Figure~\ref{fig:lnp} originates in the
  months of April-May and October-November.}\label{fig:monthly}
\end{figure}

As input we take the average CME mass in 10-day bins.  The resulting
periodogram is shown in Figure~\ref{fig:lnp}. We find a very
significant peak at a period of 179.1 days. The probability of this
peak being due to random noise is only $3.7\times 10^{-11}$. To check
for effects of binning, we repeated the analysis with different
temporal bins without any change in the results. The peak remains at
the same period.  When we repeat the analysis on a yearly basis, we find
that the 6-month peak arises only from 2003 onwards. We did not detect
any significant peaks during the previous minimum in 1996-1999. The
mission interruption and synoptic program changes during 1996-1997 may
have played a role. We also searched for possible instrumental causes
for this variability.  It is, however, difficult to see how such a
long-term periodicity can arise when we are dealing with essentially
base-difference images taken a few hours apart.  Spacecraft operations
can also be dismissed since they occur on a 3-month basis and are
subtracted out in any case when the pre-event image is subtracted.  To
find out when the activity peaks during the year, we calculated the
average CME mass per calendar month for the whole database in
figure~\ref{fig:monthly}.  The 6-month variability is obvious in the
plot and corresponds to the months of April-May and
October-November. Indeed, these months are related to increased solar
activity since the beginning of the SOHO mission (e.g., see numerous
papers on November 1997, April-May 1998, October-November 2003
events).  

This is the not the first time that a 6-month periodicity is
reported. \citet{2008SoPh..248..155L} found a 193-day periodicity in
a spectral analysis of CME rates from 1995 to 2006. Their sample
included all events in the CDAW list so the 13-day difference with our
results may stem from the inclusion of the smaller events or the use
of rates instead of a physical property such as mass or both. A
well-known '154-day' periodicity has been reported for many other
solar activity indicators such as flares, sunspot numbers, ICME rates,
flux emergence \citep[see][and references
therein]{2005GeoRL..3202104R}. Those studies have shown that the
periodicity is quite wide and can extend from 154 to 190 days,
depending on the analysis method. In our case, we could only detect
the 180-day periodicity and there were no peaks in other periods
(Figure~\ref{fig:lnp}.  It may be that such periodicities change from
cycle to cycle as suggested by \citet{2005GeoRL..3202104R}.  In
addition, there has not been a satisfactory explanation for the
154-day periodicity in the other solar parameters nor can we provide
one for the 180-day periodicity in our data.  We can only point out
that our periodicity is directly related to the appearance of large
active region complexes on the sun and therefore must be related to
processes that control magnetic flux emergence from the convection
zone. A similar conclusion was reached by \citet{2008SoPh..248..155L}.
More studies are surely necessary and may be able to provide new
insights into the link between flux emergence and the CME ejection
phenomenon.
 
We close this section by noting that starting in 2006, the monthly
ejection rate shows less fluctuations but the 180-day periodicity
persists. The rate has a slower downward trend and reaches activity
levels similar to 1996-1997. In other words, the minima of cycles 22
and 23 are similar. The last point in this graph is unusually low
because it contains only a third of the number of events (8 CMEs) in
any of the previous points ($>20$).

\section{Conclusions}
In the preceding sections, we presented an extensive analysis of the
first full solar cycle database of CMEs from the viewpoint of their
mass and energy properties. This work provided us with the opportunity
to describe in detail the mass measurement procedures with their error
analyses for the first time in the literature. Our measurements are
incorporated in the online database of the LASCO PI team at NRL
(\url{http://lasco-www.nrl.navy.mil/}) and our initial statistical
analyses provided new insights on the interpretation
of CME measurements and the nature of CMEs. We summarize the main
results of our study as follows:
\begin{itemize}
\item We identify the existence of two
  populations in the mass CME data: The 'normal' CMEs which reach a
  constant mass beyond 10 R$_\sun$, and the 'pseudo' CMEs which reach
  a mass peak below 7 R$_\sun$ before they disappear in the C3 field
  of view.
\item The average CME mass density is remarkably constant for the
  'normal' CMEs at $\rho = 10^{12.83\pm0.25}$
  g/R$_{\sun}^2$. Therefore, one can calculate the mass of any CME,
  within a factor of 3, simply by multiplying this density with the
  projected area of the CME, in $R_\sun^2$, as measured in the
  images. The CME volumetric density, calculated under the assumption
  of line-of-sight width equal to the observed width, is $\sim3500$
  cm$^{-3}$ and suggests that most of the CME mass originates in the
  high corona, if CMEs expand adiabatically.
\item The mass and energy distributions become log-normal only for
  measurements at around 10-15 R$_\sun$. The implication is that
  measurements at lower heights provide an incomplete picture of the
  event and will bias statistical studies. Coronagraphs with FOVs at or
  beyond 15 R$_\sun$ are essential for CME studies. 
\item We show the first complete solar cycle behavior of the CME mass,
  density, and energies. The statistics reveal a sudden drop in the
  CME mass, mass density, and energy which takes place within two
  Carrington rotations (2005-2006) in June-August, 2003. This is
  followed by an increased monthly variability in these CME properties
  that lasts until mid-2005 with a very strong 6-month periodicity
  which continues to this day. The reason for this sudden change is
  unclear at this moment but it seems to signal a transition of the
  large scale coronal electron content towards minimum levels. The
  cause of the 6-month variability is also under investigation.
\item We show that the CME properties also vary between cycles. CMEs
  seems to be less massive (by almost a factor of 10) compared to
  cycle 21 measurements. Also, the CME properties at the current
  minimum  are below their values in the
  previous minimum by factors of 4.6 for the mass, 6.4 for the energy and only
  17\% for the speed. These observations provide additional evidence for the
  unusual minimum of cycle 23.
\item The LASCO duty cycle is $\sim94\%$.
\item Instrumental errors and assumptions on the composition of the
  plasma are insignificant compared to the projection errors below.
\item CME masses and potential energies may be underestimated by a
  maximum of 2x, and the kinetic energies by a maximum of 8x (mostly
  due to the speed projection). 
\end{itemize}

The above results constitute only a subset of the possible analyses
that can be performed with such a large dataset of CME
measurements. The current analysis has established some of the typical
properties, such as the typical CME density, mass flux, and the
distribution of CME energies and masses and has raised new questions
regarding the solar cycle variation of CME mass.  We plan to continue
updating the database while the operation of the LASCO coronagraphs
continues.

\begin{acknowledgements}
  SOHO is an international collaboration between NASA and ESA. LASCO
  was constructed by a consortium of institutions: the Naval Research
  Laboratory (Washington, DC, USA), the Max-Planck-Institut fur
  Aeronomie (Katlenburg- Lindau, Germany), the Laboratoire
  d'Astronomie Spatiale (Marseille, France) and the University of
  Birmingham (Birmingham, UK). The LASCO CME catalog is generated and
  maintained at the CDAW Data Center by NASA and The Catholic
  University of America in cooperation with the Naval Research
  Laboratory. 
\end{acknowledgements}

\bibliographystyle{apj}
\bibliography{apj-jour,cmes}

\begin{thebibliography}{50}
\expandafter\ifx\csname natexlab\endcsname\relax\def\natexlab#1{#1}\fi

\bibitem[{{Abramenko} \& {Longcope}(2005)}]{2005ApJ...619.1160A}
{Abramenko}, V.~I., \& {Longcope}, D.~W. 2005, \apj, 619, 1160

\bibitem[{{Andrews} {et~al.}(1999){Andrews}, {Wang}, \&
  {Wu}}]{1999SoPh..187..427A}
{Andrews}, M.~D., {Wang}, A., \& {Wu}, S.~T. 1999, \solphys, 187, 427

\bibitem[{{Aoki} {et~al.}(2003){Aoki}, {Yashiro}, \&
  {Shibata}}]{2003ICRC....5.2729A}
{Aoki}, S., {Yashiro}, S., \& {Shibata}, K. 2003, in International Cosmic Ray
  Conference, Vol.~5, International Cosmic Ray Conference, 2729--+

\bibitem[{{Bewsher} {et~al.}(2008){Bewsher}, {Harrison}, \&
  {Brown}}]{2008A&A...478..897B}
{Bewsher}, D., {Harrison}, R.~A., \& {Brown}, D.~S. 2008, \aap, 478, 897

\bibitem[{{Billings}(1966)}]{billings:66}
{Billings}, D.~E. 1966, {A guide to the solar corona} (New York: Academic Press
  Inc)

\bibitem[{{Brueckner} {et~al.}(1995){Brueckner}, {Howard}, {Koomen},
  {Korendyke}, {Michels}, {Moses}, {Socker}, {Dere}, {Lamy}, {Llebaria},
  {Bout}, {Schwenn}, {Simnett}, {Bedford}, \& {Eyles}}]{lasco:95}
{Brueckner}, G.~E., {et~al.} 1995, \solphys, 162, 357

\bibitem[{{Ciaravella} {et~al.}(2003){Ciaravella}, {Raymond}, {van
  Ballegooijen}, {Strachan}, {Vourlidas}, {Li}, {Chen}, \&
  {Panasyuk}}]{2003ApJ...597.1118C}
{Ciaravella}, A., {Raymond}, J.~C., {van Ballegooijen}, A., {Strachan}, L.,
  {Vourlidas}, A., {Li}, J., {Chen}, J., \& {Panasyuk}, A. 2003, \apj, 597,
  1118

\bibitem[{{Colaninno} \& {Vourlidas}(2009)}]{2009ApJ...698..852C}
{Colaninno}, R.~C., \& {Vourlidas}, A. 2009, \apj, 698, 852

\bibitem[{{Cremades} \& {St.~Cyr}(2007)}]{2007AdSpR..40.1042C}
{Cremades}, H., \& {St.~Cyr}, O.~C. 2007, Advances in Space Research, 40, 1042

\bibitem[{{Crow} \& {Shimizu}(1988)}]{crow:88}
{Crow}, E.~L., \& {Shimizu}, K. 1988, {Lognormal Distributions: Theory and
  Applications} (New York: Marcel Dekker, Inc)

\bibitem[{{Domingo} {et~al.}(1995){Domingo}, {Fleck}, \& {Poland}}]{soho:95}
{Domingo}, V., {Fleck}, B., \& {Poland}, A.~I. 1995, \solphys, 162, 1

\bibitem[{{Filippov} \& {Koutchmy}(2002)}]{2002SoPh..208..283F}
{Filippov}, B., \& {Koutchmy}, S. 2002, \solphys, 208, 283

\bibitem[{{Gilbert} {et~al.}(2001){Gilbert}, {Serex}, {Holzer}, {MacQueen}, \&
  {McIntosh}}]{2001ApJ...550.1093G}
{Gilbert}, H.~R., {Serex}, E.~C., {Holzer}, T.~E., {MacQueen}, R.~M., \&
  {McIntosh}, P.~S. 2001, \apj, 550, 1093

\bibitem[{{Gopalswamy} {et~al.}(2003){Gopalswamy}, {Lara}, {Yashiro}, \&
  {Howard}}]{2003ApJ...598L..63G}
{Gopalswamy}, N., {Lara}, A., {Yashiro}, S., \& {Howard}, R.~A. 2003, \apjl,
  598, L63

\bibitem[{{Gopalswamy} {et~al.}(2004){Gopalswamy}, {Nunes}, {Yashiro}, \&
  {Howard}}]{2004AdSpR..34..391G}
{Gopalswamy}, N., {Nunes}, S., {Yashiro}, S., \& {Howard}, R.~A. 2004, Advances
  in Space Research, 34, 391

\bibitem[{{Gopalswamy} {et~al.}(2009){Gopalswamy}, {Yashiro}, {Michalek},
  {Stenborg}, {Vourlidas}, {Freeland}, \& {Howard}}]{2009EM&P..104..295G}
{Gopalswamy}, N., {Yashiro}, S., {Michalek}, G., {Stenborg}, G., {Vourlidas},
  A., {Freeland}, S., \& {Howard}, R. 2009, Earth Moon and Planets, 104, 295

\bibitem[{{Hayes} {et~al.}(2001){Hayes}, {Vourlidas}, \& {Howard}}]{hayes:01}
{Hayes}, A.~P., {Vourlidas}, A., \& {Howard}, R.~A. 2001, \apj, 548, 1081

\bibitem[{{Hildner} {et~al.}(1975){Hildner}, {Gosling}, {MacQueen}, {Munro},
  {Poland}, \& {Ross}}]{1975SoPh...42..163H}
{Hildner}, E., {Gosling}, J.~T., {MacQueen}, R.~M., {Munro}, R.~H., {Poland},
  A.~I., \& {Ross}, C.~L. 1975, \solphys, 42, 163

\bibitem[{{Howard} {et~al.}(2009){Howard}, {Battams}, {Vourlidas}, {Morrill},
  \& {Stenborg}}]{2009AGUFMSH13C..04H}
{Howard}, R.~A., {Battams}, K., {Vourlidas}, A., {Morrill}, J.~S., \&
  {Stenborg}, G. 2009, AGU Fall Meeting Abstracts, C4+

\bibitem[{{Howard} {et~al.}(1985){Howard}, {Sheeley}, {Michels}, \&
  {Koomen}}]{russ:85}
{Howard}, R.~A., {Sheeley}, Jr., N.~R., {Michels}, D.~J., \& {Koomen}, M.~J.
  1985, \jgr, 90, 8173

\bibitem[{Hundhausen(1993)}]{hundhausen_1993}
Hundhausen, A.~J. 1993, Journal of Geophysical Research, 98, 13177

\bibitem[{{Hundhausen} {et~al.}(1994){Hundhausen}, {Stanger}, \&
  {Serbicki}}]{1994ESASP.373..409H}
{Hundhausen}, A.~J., {Stanger}, A.~L., \& {Serbicki}, S.~A. 1994, in ESA
  Special Publication, Vol. 373, Solar Dynamic Phenomena and Solar Wind
  Consequences, the Third SOHO Workshop, ed. {J.~J.~Hunt}, 409--+

\bibitem[{{Jackson} \& {Howard}(1993)}]{1993SoPh..148..359J}
{Jackson}, B.~V., \& {Howard}, R.~A. 1993, \solphys, 148, 359

\bibitem[{{Kahler} \& {Vourlidas}(2005)}]{2005JGRA..11012S01K}
{Kahler}, S.~W., \& {Vourlidas}, A. 2005, Journal of Geophysical Research
  (Space Physics), 110, 12

\bibitem[{Krall \& St.~Cyr(2006)}]{krall_st.cyr_2006}
Krall, J., \& St.~Cyr, O.~C. 2006, The Astrophysical Journal, 652, 1740

\bibitem[{{Lara} {et~al.}(2008){Lara}, {Borgazzi}, {Mendes}, {Rosa}, \&
  {Domingues}}]{2008SoPh..248..155L}
{Lara}, A., {Borgazzi}, A., {Mendes}, Jr., O., {Rosa}, R.~R., \& {Domingues},
  M.~O. 2008, \solphys, 248, 155

\bibitem[{Limpert {et~al.}(2001)Limpert, Stahel, \&
  Abbt}]{limpert_stahel_abbt_2001}
Limpert, E., Stahel, W.~A., \& Abbt, M. 2001, BioScience, 51, 341

\bibitem[{{Llebaria} {et~al.}(2006){Llebaria}, {Lamy}, \&
  {Danjard}}]{2006Icar..182..281L}
{Llebaria}, A., {Lamy}, P., \& {Danjard}, J. 2006, \icarus, 182, 281

\bibitem[{{Lomb}(1976)}]{1976Ap&SS..39..447L}
{Lomb}, N.~R. 1976, \apss, 39, 447

\bibitem[{{Lugaz} {et~al.}(2005){Lugaz}, {Manchester}, \&
  {Gombosi}}]{2005ApJ...627.1019L}
{Lugaz}, N., {Manchester}, IV, W.~B., \& {Gombosi}, T.~I. 2005, \apj, 627, 1019

\bibitem[{{MacQueen} {et~al.}(2001){MacQueen}, {Burkepile}, {Holzer},
  {Stanger}, \& {Spence}}]{2001ApJ...549.1175M}
{MacQueen}, R.~M., {Burkepile}, J.~T., {Holzer}, T.~E., {Stanger}, A.~L., \&
  {Spence}, K.~E. 2001, \apj, 549, 1175

\bibitem[{Mittal {et~al.}(2009)Mittal, Pandey, Narain, \&
  Sharma}]{mittal_properties_2009}
Mittal, N., Pandey, K., Narain, U., \& Sharma, S.~S. 2009, Astrophysics and
  Space Science, 323, 135

\bibitem[{{Morrill} {et~al.}(2006){Morrill}, {Korendyke}, {Brueckner},
  {Giovane}, {Howard}, {Koomen}, {Moses}, {Plunkett}, {Vourlidas},
  {Esfandiari}, {Rich}, {Wang}, {Thernisien}, {Lamy}, {Llebaria}, {Biesecker},
  {Michels}, {Gong}, \& {Andrews}}]{jeff:06}
{Morrill}, J.~S., {et~al.} 2006, \solphys, 233, 331

\bibitem[{{Poland} {et~al.}(1981){Poland}, {Howard}, {Koomen}, {Michels}, \&
  {Sheeley}}]{1981SoPh...69..169P}
{Poland}, A.~I., {Howard}, R.~A., {Koomen}, M.~J., {Michels}, D.~J., \&
  {Sheeley}, Jr., N.~R. 1981, \solphys, 69, 169

\bibitem[{{Reinard}(2008)}]{2008ApJ...682.1289R}
{Reinard}, A.~A. 2008, \apj, 682, 1289

\bibitem[{{Reiner} {et~al.}(2003){Reiner}, {Vourlidas}, {Cyr}, {Burkepile},
  {Howard}, {Kaiser}, {Prestage}, \& {Bougeret}}]{2003ApJ...590..533R}
{Reiner}, M.~J., {Vourlidas}, A., {Cyr}, O.~C.~S., {Burkepile}, J.~T.,
  {Howard}, R.~A., {Kaiser}, M.~L., {Prestage}, N.~P., \& {Bougeret}, J. 2003,
  \apj, 590, 533

\bibitem[{{Richardson} \& {Cane}(2005)}]{2005GeoRL..3202104R}
{Richardson}, I.~G., \& {Cane}, H.~V. 2005, \grl, 32, 2104

\bibitem[{{Robbrecht} {et~al.}(2009){Robbrecht}, {Patsourakos}, \&
  {Vourlidas}}]{2009ApJ...701..283R}
{Robbrecht}, E., {Patsourakos}, S., \& {Vourlidas}, A. 2009, \apj, 701, 283

\bibitem[{{Scargle}(1982)}]{1982ApJ...263..835S}
{Scargle}, J.~D. 1982, \apj, 263, 835

\bibitem[{{Sheeley} {et~al.}(1997){Sheeley}, {Wang}, {Hawley}, {Brueckner},
  {Dere}, {Howard}, {Koomen}, {Korendyke}, {Michels}, {Paswaters}, {Socker},
  {St.~Cyr}, {Wang}, {Lamy}, {Llebaria}, {Schwenn}, {Simnett}, {Plunkett}, \&
  {Biesecker}}]{1997ApJ...484..472S}
{Sheeley}, Jr., N.~R., {et~al.} 1997, \apj, 484, 472

\bibitem[{{St.~Cyr} {et~al.}(2000){St.~Cyr}, {Plunkett}, {Michels},
  {Paswaters}, {Koomen}, {Simnett}, {Thompson}, {Gurman}, {Schwenn}, {Webb},
  {Hildner}, \& {Lamy}}]{2000JGR...10518169S}
{St.~Cyr}, O.~C., {et~al.} 2000, \jgr, 105, 18169

\bibitem[{Stewart {et~al.}(1974)Stewart, McCabe, Koomen, Hansen, \&
  Dulk}]{stewart_mccabe_koomen_hansen_dulk_1974}
Stewart, R.~T., McCabe, M.~K., Koomen, M.~J., Hansen, R.~T., \& Dulk, G.~A.
  1974, Solar Physics, 36, 203

\bibitem[{{Subramanian} \& {Vourlidas}(2007)}]{prasad:07}
{Subramanian}, P., \& {Vourlidas}, A. 2007, \aap, 467, 685

\bibitem[{{Thernisien} {et~al.}(2009){Thernisien}, {Vourlidas}, \&
  {Howard}}]{2009SoPh..256..111T}
{Thernisien}, A., {Vourlidas}, A., \& {Howard}, R.~A. 2009, \solphys, 256, 111

\bibitem[{{Vourlidas} {et~al.}(2002){Vourlidas}, {Buzasi}, {Howard}, \&
  {Esfandiari}}]{vour:02}
{Vourlidas}, A., {Buzasi}, D., {Howard}, R.~A., \& {Esfandiari}, E. 2002, in
  ESA Special Publication, Vol. 506, Solar Variability: From Core to Outer
  Frontiers, ed. {J.~Kuijpers}, 91--94

\bibitem[{{Vourlidas} \& {Howard}(2006)}]{v_h:06}
{Vourlidas}, A., \& {Howard}, R.~A. 2006, \apj, 642, 1216

\bibitem[{{Vourlidas} {et~al.}(2000){Vourlidas}, {Subramanian}, {Dere}, \&
  {Howard}}]{vour:00}
{Vourlidas}, A., {Subramanian}, P., {Dere}, K.~P., \& {Howard}, R.~A. 2000,
  \apj, 534, 456

\bibitem[{{Yashiro} {et~al.}(2003){Yashiro}, {Gopalswamy}, {Michalek}, \&
  {Howard}}]{2003AdSpR..32.2631Y}
{Yashiro}, S., {Gopalswamy}, N., {Michalek}, G., \& {Howard}, R.~A. 2003,
  Advances in Space Research, 32, 2631

\bibitem[{{Yashiro} {et~al.}(2004){Yashiro}, {Gopalswamy}, {Michalek},
  {St.~Cyr}, {Plunkett}, {Rich}, \& {Howard}}]{2004JGRA..10907105Y}
{Yashiro}, S., {Gopalswamy}, N., {Michalek}, G., {St.~Cyr}, O.~C., {Plunkett},
  S.~P., {Rich}, N.~B., \& {Howard}, R.~A. 2004, Journal of Geophysical
  Research (Space Physics), 109, 7105

\bibitem[{{Yurchyshyn} {et~al.}(2005){Yurchyshyn}, {Yashiro}, {Abramenko},
  {Wang}, \& {Gopalswamy}}]{2005ApJ...619..599Y}
{Yurchyshyn}, V., {Yashiro}, S., {Abramenko}, V., {Wang}, H., \& {Gopalswamy},
  N. 2005, \apj, 619, 599

\end{thebibliography}

\begin{deluxetable}{lcl}
\tablecolumns{3}
\tablewidth{0pc}
\tablecaption{Typical Errors in the Calculation of CME Mass} \label{tab:dn2msb}
\tablehead{
\colhead{Operation} & \colhead{Parameter} & \colhead{$\sigma_x/x (\%)$}}
\startdata
DN $\rightarrow$ MSB: $I = {{S}\over{t}}*V*F_{cal}$  & I   & $1.98-2.3$ \\
 & S   & 0.4 - 1.4 \\ 
 & t   & 0.15 \\ 
 & V   & 1 \\ 
 &F$_{cal}$ & 0.73 \\[0.1pc] 
\hline
MSB $\rightarrow$ excess MSB: $I_{CME} = I - I_{pre}$ & $I_{CME}$ & 4\tablenotemark{a}\\
 & $I_{CME}$ & $0.005$\tablenotemark{b}\\[0.1pc]
\hline
Excess MSB $\rightarrow$ Mass: $M_{CME} = I_{CME}*C_e*C_{plasma}$ & $M_{CME}$ & 100.2\\
 & $C_e$ &  -100\\
 & $C_{plasma}$ & -6\\
\hline
\enddata
\tablenotetext{a}{Assuming independent variables}
\tablenotetext{b}{Based on C3 data (Figure~\ref{fig:excess_stat})}
\end{deluxetable}
\begin{deluxetable}{lcccc}
\tablecolumns{5}
\tablewidth{0pc}
\tablecaption{Mass and Energy Properties of CMEs
  (1996-2009)}\label{tbl:properties}
\tablehead{
\colhead{}  &\multicolumn{3}{c}{LASCO} & \colhead{Solwind} \\
\cline{2-5} \\
\colhead{Property} & \colhead{Histogram Peak} & \colhead{Average} &
  \colhead{Median} & \colhead{Average}}
\startdata
Mass ($\times 10^{14}$ g)         & 3.4 & 3.9   & 11    & 17\\
$E_K$ ($\times 10^{29}$ ergs)      & 3.4 & 2.3   & 19     & 43\\
$E_{mech}$ ($\times 10^{29}$ ergs)  & 8.5 & 9.0   & 38     & \nodata\\
\hline
Total Mass ($\times10^{18}$ g)   &\nodata &\nodata& 9.2  & 3.9\\
Mass Flux ($\times 10^{15}$ g/day)&\nodata &\nodata&
1.8\tablenotemark{a}  & 7.5\\  
Duty Cycle                       &\nodata &\nodata& 94\% & 61.7\% \\
\enddata 
\tablenotetext{a}{1996-2003: $2.6\times10^{15}$ g/day. 2003-2009: $8.3\times10^{14}$ g/day.}
\end{deluxetable}
\begin{deluxetable}{lccc}
\tablecolumns{5}
\tablewidth{0pc}
\tablecaption{Results of the Normal Fit to the log of Mass and Energy}\label{tbl:logfit}
\tablehead{
\colhead{Parameter}  &\colhead{$\mu$} & \colhead{Geometric Mean ($e^\mu$)} & \colhead{$\sigma$}
}
\startdata
Mass  (g)        & 34.974 & $1.55\times10^{15}$  & 1.114   \\
$E_K$ (ergs)      & 69.748 & $1.96\times10^{30}$ & 1.515   \\
$E_{mech}$  (ergs) & 70.682 & $4.98\times10^{30}$  & 1.184 \\
\enddata 
\end{deluxetable}
\clearpage
\end{document}